\documentclass[a4paper,12pt]{article}

\usepackage{latexsym}
\usepackage{amsmath}
\usepackage{amssymb}
\usepackage{enumitem}

\begin{document}

\title{Test--field limit of metric nonlinear gravity theories}

\author{Guido MAGNANO \\
\small Dipartimento di Matematica ``G.~Peano'',
Universit\`a di Torino, \\
\small via Carlo Alberto 10, 10123 Torino, Italy \\[5mm]
Krzysztof A. MEISSNER \\
\small Faculty of Physics, University of Warsaw\\
\small Pasteura 5, 02-093, Warsaw, Poland\\[5mm]
Leszek M. SOKO\L{}OWSKI \\
\small Astronomical Observatory,
Jagellonian University\\
\small Orla
171, 30-244 Krak\'ow, Poland\\
\small and Copernicus Center for Interdisciplinary Studies, Krak\'ow, Poland}
\date{}
\maketitle

\begin{abstract}
In the framework of alternative metric gravity theories, it has been shown by several authors that a generic Lagrangian depending on the 
Riemann tensor describes a theory with 8 degrees of freedom (which reduce to 3 for $f(R)$ Lagrangians depending only on the curvature scalar).
This result is often related to a reformulation of the fourth-order equations for the metric into a set of second-order equations for a multiplet of fields, including -- besides the metric -- a massive scalar field and a massive spin-2 field (the latter being usually regarded as a ghost): this is commonly assumed to represent the particle spectrum of the theory. In this article we investigate an issue which does not seem to have been addressed so far: 
in ordinary general-relativistic field theories, all fundamental fields (i.e.~fields with definite spin and mass) reduce to \emph{test fields} in some appropriate limit of the model, where they cease to act as sources for the metric curvature. 
In this limit, each of the fundamental fields can be excited from its ground state \emph{independently from the others} (which does not happen, instead, as long as the fields are coupled through the gravitational interaction).
The question is: does higher-derivative gravity admit a test-field limit for its fundamental fields?
It is easy to show that for a generic $f(R)$ theory the test-field limit does exist; then, we consider the case of Lagrangians depending on the full Ricci tensor. We show that, already for a quadratic Lagrangian, the constraint binding together the scalar field and the massive spin-2 field does not disappear in the limit where they should be expected to act as test fields. A proper test-field limit exists only for a particular choice of the coefficients in the Lagrangian, which cause the scalar field to disappear (so that the resulting model has only 7 d.o.f.). 
We finally consider the possible addition of an arbitrary function of the quadratic invariant of the Weyl tensor, $C^{\alpha}{}_{\beta\mu\nu}C_{\alpha}{}^{\beta\mu\nu}$, and show that after this addition the resulting model still lacks a proper test-field limit. We argue that the lack of a test-field limit for the dynamics of the fundamental fields may constitute a serious drawback of the full 8 d.o.f.~higher-order gravity models, which is not encountered in the restricted 7 d.o.f.~or 3 d.o.f.~cases.
\end{abstract}

\section{Introduction}
Alternative theories of gravity have appeared soon after the advent of general 
relativity, the first of them was Weyl's theory (1919) and since then they have been 
copiously created. 

This phenomenon on one hand resembles the situation in other branches of physics, for example particle physics, but on the other is very different. In particle physics, after the discovery of the Higgs boson, there is no single direct experimental evidence that the Standard Model (SM) should be enlarged or modified. However, we know that it should: first, because there are indirect observational hints from cosmology like matter-antimatter asymmetry in the Universe and second, theoretical extrapolation of SM to high energies shows pathologies in the pure SM like instabilities or Landau poles. Therefore there are many proposals of extensions of SM, either within particle physics, mostly based on supersymmetry (or conformal symmetry), or more radical, like string theory, that were designed to cure these problems. Within particle physics, under the assumption of renormalizability, our choice of possibilities is limited to different gauge groups or different field content; within string theory the fundamental theory is essentially unique but through compactification we can have an almost continuous manifold of effective low-energy 4-dimensional theories. In either case we know that SM should be extended.

In gravitational physics the situation is different. From the experimental point of view we also don't need any competing theory to Einstein's GR because of lack of any experimental evidence pointing to a necessity of supplementing the Einstein-Hilbert action with higher order terms. However, reasoning behind any modification of the simplest GR is not as convincing and straightforward as in particle physics for two reasons. First, higher order terms in Einstein's theory can be relevant only at gravitational curvatures many orders of magnitude bigger than what we could imagine as even indirectly observable in any foreseeable future (inside black holes or in the very early Universe). Second, although there are hints that such terms are necessary in quantum theory of fields coupled to gravity we don't have quantum theory of gravity itself to make such a picture fully consistent. 

The theoretical arguments pointing to a necessity of adding higher order terms in gravity come from two sources. The first one is string theory or (not yet fully constructed) more general UV completion of the theory of particles, called M theory. The theory of closed strings has a graviton as its massless excitation. We can calculate $N$-point amplitudes involving gravitons as external legs and these amplitudes can be translated as coming from the effective theory of gravity with higher order terms in the Riemann curvature expanded around the flat geometry. There is however another approach leading directly to these corrections without any expansion: the calculation of the (2-dimensional) conformal anomaly of string propagating on a curved background. Then the requirement of vanishing of this anomaly leads directly to the Einstein equations with higher order corrections \cite{GSW}.

The second argument that points to higher order corrections in GR comes from the theory of particles coupled to gravity without invoking any UV completion. Then the loop calculations with massive particles on internal lines and gravitons as external legs lead at low curvatures to local higher order corrections in the Riemann tensor. These calculations produce infinities (in contrast to string theory where amplitudes are finite due to modular invariance) and require renormalization. Unfortunately, gravity is non-renormalizable and at each level we have to introduce a new renormalization prescription what makes such a theory useless from the physical point of view. Another issue is a ($D$-dimensional) conformal anomaly that also contains higher powers of the Riemann tensor but even at low curvatures the anomaly is non-local and therefore has to vanish because of potentially disastrous consequences for observable theory of gravity \cite{MN1}. 

All these arguments point to the fact that any quantum theory of particles coupled to gravity leads to an effective theory of gravity with higher order corrections. All effective theories have a limited application range -- they are good approximations only in certain interval of energies. In this case (barring a possibility of conformal anomaly which is non-local even in the IR regime) the limiting energy is presumably close to the Planck scale where quantum gravity effects start to be important and/or more massive particles should be added. It seems worthwhile to consider consequences of adding higher order corrections below the Planck scale, i.e.~at a purely classical level. For the purpose of this paper it is important to emphasize that the effective theories even with higher order terms or non-localities remain {\it bona fide} field theories (either classical or quantum).  Therefore, assuming that there exists a quantum version of these theories, in some appropriate limit of coupling constants approaching 0 these theories should admit an interpretation in terms of masses and spins (except in 2 dimensions where there may exist some 'pathologies' associated with the fact that the little group has continuous representations). Even if the theory admits for non-vanishing coupling constant $g$ non-trivial classical solutions, like solitons or instantons, that do not allow to take the limit $g\to 0$, in the quantum interpretation they are not treated as separate particles but as an (infinite) collection of excitations out of a vacuum that are saddle points in the path integral (either Euclidean or Lorentzian).  

In this paper we are dealing with one class of alternative theories, which differ from 
GR in only one axiom: in gravitational field equations. Instead of Einstein--Hilbert 
Lagrangian, $L=R$, one assumes that $L$ is some smooth scalar function of the Riemann 
tensor. These theories have no commonly accepted name and they are dubbed either as 
`metric nonlinear gravity' (NLG, Lagrangians nonlinear in the curvature) or `higher 
derivative theory' (though the field equations need not be of fourth order); we 
shall use the first term. A revival of these theories has taken place after the 
discovery in 1999 of the apparent accelerated evolution of the universe. Most of the 
researchers attempting to account for this acceleration without resorting to the 
mysterious dark energy or the fine tuned cosmological constant, have applied various 
$f(R)$ gravity theories (see \cite{DFT} for a review). However, once one rejects the 
simplest Lagrangian $L=R$,  
one is not restricted to a function $f(R)$ of the curvature scalar, and one may 
investigate the generic case of $L=f(g_{\mu\nu}, R^{\alpha}{}_{\beta\mu\nu})$ and 
pass over any dependence on derivatives of the tensor. Hence, prior to any cosmological 
applications of any NLG theory one should firmly establish on physical grounds (by 
theoretical arguments since the observational data are insufficient) the correct 
Lagrangian. From a mathematical 
viewpoint there is nothing inappropriate in assuming that $L$ depends on $R$, the 
Ricci tensor $R_{\mu\nu}$ and the Weyl tensor $C^{\alpha}{}_{\beta\mu\nu}$ apart from 
that the field equations become formidably complicated in comparison to Einstein's 
field equations. Yet from a physical viewpoint severe doubts arise. All metric 
nonlinear gravity theories are inherently ambiguous in their physical 
interpretation due to possibility of performing various redefinitions of
their dynamical variables and only Einstein's GR is free of these ambiguities: 
this fact was gradually discovered in a series of works (see \cite{K2,SoFar,NGLR}  for full references). 

It has been shown by various authors (starting with Stelle \cite{KS}) that a Lagrangian depending on the Ricci tensor of a metric -- in dimension four and without external matter -- corresponds, generically, to a model with eight degrees of freedom (in the field--theoretical sense). For a quadratic Lagrangian, the inclusion of the Weyl tensor does not increase the number of DOF. In the case of a (generically nonlinear) Lagrangian $f(R)$ depending only on the curvature scalar, the DOF reduce to three. According to common wisdom, a Lagrangian depending in a generic way on the Ricci tensor is equivalent to a second--order field theory including a massless spin-two field, a massive spin-two field and a massive scalar field. In the $f(R)$ case the massive spin-two field is absent.

Apart from the $f(R)$ case, nearly all theories with higher order terms in the Riemann tensor suffer from the presence of ghosts, i.e. fields with negative 
coefficients in front of the kinetic terms\footnote{Among gravitational theories with higher order terms there is a class of theories 
that is free from ghosts: the Lovelock theories, based on topological forms. }. Such a situation is quite common in gauge theories where these fields are associated with gauge degrees of freedom. 
There are two approaches to this problem in quantum theory: either we gauge away these fields (what usually results in a much more complicated theory, since 
we explicitly break the symmetry to keep only positive kinetic terms), or we allow for quantization of these terms as well but later we decouple quantum 
states associated to these fields from the rest (Gupta-Bleuler quantization). The presence of ghosts is expected to lead to vacuum instability already at 
the classical level. However, it has been shown in \cite{MS2} that for Lagrangian depending quadratically on the Ricci tensor the ground--state solution, 
where the massive spin--two field is set to zero, is linearly stable against small excitations of this field. The consequences of the ghost nature of the 
massive spin--two component of the multiplet are thus still an open question, but we shall not deal with this problem in this article; we shall only make 
a short comment on derivation of this property at the end of Sect. 4. 

We shall instead focus on a different problem arising when one tries to 
identify NLG theories with second--order field theories involving a multiplet of fields, or -- in a quantum perspective --  to attach particle degrees of 
freedom to higher-derivative terms.

As we have already recalled, in standard field theory each fundamental field is expected to have a definite mass and spin: 
if definite values of these quantities cannot be assigned to the field under 
consideration, this simply means that the field is actually a unifying field (a 
kind of `mixture') for a number of distinct fundamental fields.\\ 
In the case of a NLG theory, this means that the metric appearing in its Lagrangian 
is just a unifying field and should be decomposed in a multiplet of gravitational 
fields of definite spin, that we have mentioned above. The physical metric of the spacetime will be only one element of the multiplet (and need not 
be identical to the original unifying metric). We have mentioned above this decomposition; for $f(R)$ gravity theories the 
decomposition into a metric and a scalar field was performed in \cite{MFF1, JK1, JK2, 
MS1, ABJT}. The case of the quadratic Lagrangian (without the Weyl tensor), 
$L=\kappa R+aR^2+bR_{\mu\nu}R^{\mu\nu}$, is also relatively well understood \cite{MS2} (for a completely different approach to quadratic theory see 
\cite{AGKKLR}). From the physical viewpoint, the decomposition of the unifying metric into two or three 
component fields having definite masses and spins means establishing the particle 
content of the given NLG theory. 

However, this decomposition does not ensure by itself that each field of the multiplet can be regarded as an independent physical entity. In ordinary field theories, in fact, it is (tacitly) assumed that physically distinct fields can, in principle, be excited independently from each other. We write ``in principle'', because if two fields are mutually interacting then an excitation of either of them indeed produces, in general, an excitation of the other: for instance, a charged matterfield cannot be excited without affecting the state of the electromagnetic field. Nevertheless, we can claim that the two fields are physically distinct -- i.e. not merely two components of a single physical entity -- because, taking the appropriate limit of the model parameters (in the easiest situations, by setting some coupling constant equal to zero), we can produce a consistent version of the model in which both fields can live independently from each other.

When gravitation is involved the situation becomes more subtle, because any field entering the stress--energy tensor, when excited, may affect the spacetime metric and therefore alter the state of all other fields, even in the absence of direct coupling. Thus, as we discuss below in greater detail, in this case the physical content of the model should be investigated in a limit where both mutual interactions of fields and their interactions with the spacetime metric are suppressed (notice that we should not expect, in this limit, that all fields become free fields: there is no reason, in fact, to require that the \emph{self--interaction} of each field be suppressed and in some cases it is not). 

If it turns out that even in this limit it is impossible to get rid of constraints binding together the fundamental fields, then one should conclude that the theory cannot be consistently regarded as describing a multiplet of physically distinct fields.

It has been already observed in \cite{MS2} that for a quadratic Lagrangian depending on the Ricci tensor the massive spin--two field and the scalar field cannot be excited independently. Here we address the question for a larger class of models, whereby a term nonlinearly depending on the quadratic invariant of the Weyl tensor is added to the Lagrangian. In section 2 we discuss the general setting of the problem. 

NLG theories do not comprise an inherent 
method of decomposing the unifying metric into a gravitational multiplet of fields and in 
section 3 we discuss the best known decomposition procedure, based on a kind of generalized 
Legendre transformation. The procedure must be covariant and should not be confused with the (3+1)
canonical formalism applied in general relativity; actually, we employ a far going 
generalization of known Legendre transformation to obtain second-order field equations (and in consequence our formalism cannot be 
applied to GR itself). 

As a test of viability of the limit procedure, we apply it in section 4 to the quadratic Lagrangian $L=\kappa R+aR^2+bR_{\mu\nu}R^{\mu\nu}$. We stress that the 
procedure does not consist in expanding equations around the flat space solution: no truncation or linearization whatsoever is done. Instead, we push the 
three parameters entering the Lagrangian to a limit that decouples the two massive fields from the metric (this limit is singular for the Lagrangian, but not 
for the field equations). In this limit, the metric should be Ricci--flat, as a consequence of the Einstein equation for the model. The (full) equations for 
the massive fields holding in this limit show that, although the two field together carry six DOF as expected, the scalar field cannot be excited independently 
of the massive spin--two field. Hence, one cannot say that this NLG model is equivalent to a theory where five DOF are carried by a massive spin--two field 
and one DOF is carried by a massive scalar field. We shall see that this 
procedure actually comprises performing successively two independent operations to the multiplet.
 \\ 
The subsequent sections are devoted to 
Lagrangians explicitly depending on the Weyl tensor. For the fairly general form of the Weyl contribution that we are considering one can introduce,
besides the tensorial momentum conjugate to the Weyl tensor itself, a scalar momentum which subsumes the nonlinear dependence on the quadratic invariant
$C^{\alpha}{}_{\beta\mu\nu}C_{\alpha}{}^{\beta\mu\nu}$. The procedure also depends on whether the Lagrangian 
explicitly contains the Ricci tensor or not. Accordingly, sections 5 to 8 consider the 
corresponding four cases. \\
The common outcome of these cases is that the presence of the conformal tensor 
makes the interpretation of the theory in terms of a multiplet of fields (particles) 
obscure or even impossible. The paper is focused on concepts, methods and results whereas all 
calculations are reduced to the essential minimum.  

\section{Physical fields, degrees of freedom and flat spacetime}
The first assumption that we make is that any tenable theory of gravitation is a specific field 
theory. This means that most concepts and notions developed in standard classical field theory 
in flat spacetime should adequately apply (except for energy!) to gravitation. This is the case of 
Einstein's GR which can be well approximated by a linear spin--two field in Minkowski space (there 
are also attempts to formulate GR as a nonlinear field in flat space).  
We therefore postulate that any metric gravity theory should obey the tenet of Lagrangian 
field theory (both classical and quantum): a fundamental field should have a 
definite (nonnegative) mass and spin. If a field appearing in the Lagrangian (and 
in this sense a fundamental one \cite{Weinberg}) does not satisfy this condition, it
should be decomposed into a multiplet of interacting fundamental fields. Yet if 
gravitation (NLG theory) does not obey this tenet, then very little can be asserted about its 
physical interpretation.

Properties of fundamental fields (particularly masses and spins) must be determined in the coupling-free limit. 
By this expression we mean the following: in any field theory the fundamental fields can be interacting with 
each other, but if they are to be regarded as being physically distinguishable (and not just an artificial 
representation of different degrees of freedom of a single physical entity), then it should be possible to 
remove their mutual interaction, by taking a suitable limit of the coupling constants which appear in the Lagrangian. 

In this limit, the fundamental fields should behave as independent test fields: in particular, it should be possible 
to excite any one of them while the others remain in their ground state. This should be true for the full model, 
and not only in the linear approximation. We remark that in the coupling-free limit each fundamental field can 
still have a nontrivial self-interaction: it is the mutual interaction between different fields that should disappear. 
As we shall see, in our case equations of motion for the fundamental fields become linear in this limit, however one 
should not assume it in advance.

If a system of interacting fields is considered in flat spacetime, studying the coupling-free limit is relatively 
easy. Suppose that fields $\phi_1$, $\phi_2$, $\ldots\phi_n$ interact. In many cases one gets $\phi_1$ free 
just by assuming that all other fields are in their ground states, $\phi_i=0$ or $\phi_i$ are 
constant tensors for $i=2,\ldots n$. Then all equations of motion for the system (including possible 
constraints) are reduced to some equations for $\phi_1$ and one determines the properties of the field 
in the standard way; next one repeats this procedure for all other fields of the system. 

It may happen, however, that even after `switching off' all interactions by canceling their corresponding 
coupling constants, some fields in the system, say $\phi_1$ and $\phi_2$, are so strictly coupled that $\phi_1$ 
cannot be excited if $\phi_2$ is in its ground state. This may occur, for instance, due to a constraint relating 
$\phi_1$ to $\phi_2$. We then say that $\phi_1$ and $\phi_2$ are not independent physical fields. We shall show 
that this case does occur in most NLG theories, particularly if their Lagrangian contains a Weyl tensor term. 
According to our viewpoint, in these cases the original fourth-order theory \emph{cannot} be legitimately regarded 
as being equivalent to a physical second-order field theory: equivalence holds only provided the Lagrangian is 
restricted in such a way that the degrees of freedom are reduced to those carried by physically independent 
fields. We shall say that such restricted Lagrangian represents a physically tenable theory.

This procedure of identifying a dependent field may be equivalently expressed in terms of number of 
degrees of freedom (DOF). Counting of physical degrees of freedom in a theory with higher order terms in gravity 
is given for example in a seminal paper on conformal supergravity (i.e. with the Weyl squared term  in the bosonic sector) 
\cite{KTN}. The counting is quoted as an argument for the equal number of bosonic and fermionic degrees of 
freedom but 
it illustrates the main point advocated in the present paper for a general theory with higher order terms in the 
Riemann tensor. It should be emphasized that in supergravity the couplings cannot be arbitrarily put to 0 since 
the local supersymmetry provides strong constraints on such a procedure but still the counting works as 
described below.

The standard way of determining the number $F$ of DOF carried by the system 
is by applying the Hamiltonian formalism and its power lies 
in that it not only counts the degrees of freedom, but that it first of all identifies 
them. However in the case of nonlinear field systems the formalism is intricate (it is more  
intricate for NLG theories, see \cite{DSSY}, than the approach developed here) and for our 
purposes explicit identification of DOF is unnecessary; what we actually need is their number $F$. 
To this end one computes the number of independent initial data on a Cauchy surface, which is equal 
to the number of algebraically independent components of all the fields multiplied by the order of 
the Lagrange equations minus the number of all constraints imposed on these equations and following from 
them (as we recall in the next section, these include the conservation laws originating from coordinate 
invariance); 
$F$ is one half of this number \cite{PR, IN}. This approach is non--algorithmic (whereas the canonical Hamiltonian 
formalism is algorithmic, but in our case is too complex to be useful) because there is no general method of 
checking 
if all constraints have already been found; some constraints may be hard to recognize. For simple 
gravitational Lagrangians we study here this approach works effectively; moreover due to decomposition of 
gravitation into a multiplet of fields, some DOF may be precisely identified. Then one determines the 
number of DOF carried by each field of the system. To this end, as above, 
one first ensures that all the fields may be decoupled by canceling their interactions and this may require 
taking appropriate limits in all coupling constants. Next one takes one field of the system, assumes that 
all others are in their ground state and studies the resulting equation of motion for it. 
This equation together with possible constraints determines the number of DOF for the field.
Repeating this procedure for all other fields one gets the total number of DOF of fundamental fields of 
the system. 

This analysis should be done in flat spacetime, for two reasons. Firstly, physical properties of ordinary fields 
-- to which one should compare the properties of the gravitational multiplet -- are currently formulated in the 
framework of special relativity, i.e.~in Minkowski space. Secondly, gravitational interaction of a field (even if it 
is a test field) may \textit{increase\/} the number of its DOF (usually it is believed that an 
interaction may decrease the number of DOF, yet curvature may do the opposite) \cite{IN}. This may 
occur for simple Lagrangians for vector fields 
since spacetime curvature may cancel some of the constraints satisfied by the field in flat 
spacetime. In any NLG theory the corresponding gravitational multiplet, whatever way it has been 
defined (we discuss this issue in the next section), is given together with its Lagrange equations 
in a curved spacetime and making a transition to the flat one is a subtle problem. In the framework of 
GR one recovers the standard physics as 
follows. Clearly one cannot merely replace $g_{\mu\nu}$ by Minkowski metric $\eta_{\mu\nu}$ in 
Einstein field equations (EFE) because they then imply $T_{\mu\nu}=0$. Physical fields have 
positive energy density and flat spacetime trivially implies that it is empty. Instead one must 
`trivialize' EFE by breaking the coupling of any matter to geometry by taking the limit $G\rightarrow 
0$; this mathematically makes all matter a test one. Gravitation (curvature) is then free, but one 
cannot assume that the spacetime is any fixed solution of $R_{\mu\nu}=0$ or any other fixed metric. 
This is so because in a fixed but curved spacetime it is impossible to recover the standard physics,
e.g.~the integral energy and momentum conservation laws for continuous matter do not hold; 
moreover physics in a plane gravitational wave spacetime or in anti--de Sitter space is bizarre \cite{S4}. 
Gravitation may be also switched off in the framework of GR by two other methods in the full Lagrangian: 
i) by putting $R=0$ and replacing $g_{\mu\nu}$ by $\eta_{\mu\nu}$ in the matter Lagrangian, ii) since the 
gravitational Lagrangian is $c^3 R(16\pi G)^{-1}$, one can take the limit $1/G\rightarrow 0$ and 
insert the flat metric. Clearly these two methods cannot be applied to NLG theories, where one is 
interested in the gravitational multiplet and introducing ordinary matter is a subtle problem (which 
in most papers is circumvented by an arbitrary assumption) and merely makes the picture more 
intricate. 

We therefore investigate pure gravity with $L=f(g_{\mu\nu}, R^{\alpha}{}_{\beta\mu\nu})$; this Lagrangian 
must depend on at least two dimensional constants (besides $c$). Take for instance an analytic 
function $L=f(R)=\sum_n q_n R^n$. Dimensionality of any Lagrangian is $ML^{-2}T^{-1}$ ($M$---
mass, $L$---length and $T$---time), hence dimensionality of the coefficients is $[q_n]=ML^{2n-2}
T^{-1}=[c^3/G]L^{2n-2}$ and one needs one additional independent constant of dimensionality of 
length. Contributions of the Ricci and Weyl tensors to the Lagrangian may require further 
dimensional constants. All these coupling constants appear in the equations of motion for the 
gravitational multiplet of fields (these equations are equivalent to the fourth--order field 
Lagrange equations for the unifying metric $g_{\mu\nu}$). For both mathematical simplicity and 
physical interpretation the multiplet fields are so defined as to make their equations of motion 
second--order ones; in particular, the equations contain the Riemann tensor but are free of its derivatives 
and in this sense they are analogous to Einstein field equation (EFE) in GR. The gravitational multiplet consists of a 
spacetime metric (identical or not to the unifying $g_{\mu\nu}$) and a number of non--geometric 
fields\footnote{We do not know a short adequate name for these fields.} which act in these 
equations as sources for this metric. Crucial for the issue of 
tenability of the Lagrangian are properties of these non--geometric components of the multiplet 
and these features must be studied in flat spacetime, what means that one forgets their geometrical 
origin (as they arise from $L$ depending solely on $g_{\mu\nu}$) and regards them as ordinary fields consistently 
defined in Minkowski space. As in the case of GR, one cannot simply insert flat metric into the 
equations of motion since this makes them trivial: the equations of motion ``remember'' 
the geometric origin of these fields and enforce them to remain in the ground state if the metric 
is flat. The transition to flat spacetime requires 
a procedure of breaking all couplings of these fields to the curvature. In GR the coupling is 
only in EFE and to break it, it is sufficient to put $G\rightarrow 0$ whereas the form of 
matter propagation equations remains untouched, for example the electromagnetic field satisfies 
standard Maxwell equations. In a NLG theory, instead, the removal of the coupling affects the entire system of 
equations of motion (and some of them just disappear). This is accomplished by taking appropriate limits in all 
coupling constants. Furthermore one must assume that the constants tend to a limit at the same rate, so that 
their ratios remain finite in the limit.

To summarize, our method of determining the physical properties of a gravitational multiplet of fields\footnote{Clearly 
the method need not be restricted to Lagrangians considered in the present work.} comprises 
two operations which are independent, consistent and unique in providing the required outcome if performed in the 
following order.\begin{enumerate}
\item[i)] 
For the coupling constants of the multiplet fields (originally being essential dimensional parameters in the 
Lagrangian for the unifying metric) one takes the limits such that the multiplet fields cease to 
act as sources in all equations involving the Riemann tensor of the spacetime metric. 
In the model resulting from this limit, the multiplet fields (other than the metric itself) become devoid of geometrical interpretation and propagate as test fields 
in a fixed spacetime.
\item[ii)] The physical nature of these fields is investigated upon choosing, in particular, Minkowski spacetime (to avoid bizarre 
physical effects occurring e.g.~in anti--de Sitter space). This amounts to merely identifying the fixed metric with 
the flat one and then studying equations of motion.\end{enumerate}

To avoid any misinterpretation we stress once more that the method is exact, there are no approximations.

Up to now these are qualitative principles: to make them precise one should introduce a 
concrete decomposition procedure and make some simplifying assumptions on the Lagrangian. 

\section{Decomposition, various generalized Legendre transformations and the Weyl tensor}
The fourth--order equations arising from any NLG Lagrangian clearly show that one cannot assign 
a definite mass and spin to the metric appearing in it\footnote{In GR, instead, one takes the linear 
approximation $g_{\mu\nu}=\eta_{\mu\nu}+h_{\mu\nu}$ and the equations for $h_{\mu\nu}$ show that this is 
a massless field with spin (helicity) two.}, implying that it should be decomposed into 
a multiplet of gravitational fields. In the linear approximation it was first done by Stelle \cite{KS}. 
The method of decomposition in the exact theory is not inherent to it 
and should be selected in such a way as to provide the required outcome -- a well defined collection 
of fields with definite masses and spins. The correctness of the method is verified  by counting the 
degrees of freedom and this is closely related to showing the equivalence of the equations of motion 
for the multiplet to those for the unifying field. 

In other terms, first one should compute the number of DOF for the original NLG Lagrangian including 
the ``unifying metric'' alone (and its curvature components); we shall denote this number by $F$. 
After decomposing the model into field variables of definite spin and mass, one should find that the number 
of mathematical 
DOF carried by the whole multiplet (in curved space, counting the metric itself in the multiplet) coincides with 
$F$: this is a necessary condition for the correctness of the decomposition.
Having verified this one should study individual fields of the multiplet, in flat space, to identify  fields which 
are not physically independent (in the sense described in sect. 2). If such fields 
do exist, the underlying Lagrangian should be appropriately modified or rejected.\\
By inspection of the Lagrange equations for a generic metric NLG theory with 
$L=f(g_{\mu\nu}, R_{\alpha\beta\mu\nu})$ one finds that the initial metric $g_{\mu\nu}$  
carries $F=8$ DOF (see \cite{KS, BD, FT, BCh} for quadratic Lagrangians, \cite{HOW} for a perturbative method 
and \cite{S1} for a generic proof) and the above procedure is applied. 

Here two comments on counting the number of DOF in this way are in order.\\
\begin{enumerate}
\item In general relativity, the metric carries two DOF; in vacuum, there are four constraint 
equations $G^0_{\alpha}=0$. If any matter is present these are replaced by 
$G^0_{\alpha}=T^0_{\alpha}$ and in general the energy--momentum tensor does contain 
second time derivatives (of the metric and matter components) and these equations are 
no longer constraint ones. However if the matter is in the form of a scalar, vector 
or tensor (of second rank) field, it may be shown that the second time derivatives of 
the field may be removed from $T^0_{\alpha}$ with the aid of the Lagrange equations 
and the four constraints are restored \cite{S3}. A fully generic proof of that is lacking, 
nonetheless disregarding possible very exotic forms of matter being a counterexample, 
the metric has two DOF also in the presence of sources. The case of any NLG theory is 
more subtle. Due to the inherent ambiguity of these theories concerning their 
physical interpretation (i.e.~which set, or `frame', of field variables is physically 
measurable), it is unclear whether matter should be (minimally) coupled to the 
unifying metric or to the metric appearing in an appropriately redefined multiplet of 
fundamental fields (the latter choice is advocated in \cite{MS1}). Depending on the choice, 
the matter source of gravity might be a priori different from $T^{\alpha\beta}$ (be a function 
of the tensor) and the required four constraints might disappear. Fortunately, it may 
be argued that this is not the case \cite{MS1}.
\item We emphasize that independently of the actual form of the Lagrangian, four 
equations among the Lagrange ones are always constraints. Let the fourth order 
field equations of any NLG theory be denoted $E_{\mu\nu}(g)=0$. The invariance of the 
corresponding action functional under any infinitesimal coordinate transformation 
gives rise to the strong Noether conservation law (`generalized Bianchi identity')
$\nabla_{\beta}\,E^{\beta}_{\alpha}\equiv 0$ and this implies that the components 
$E^0_{\alpha}$ involve at most third time derivatives, then $E^0_{\alpha}=0$ are 
constraint equations. 
\end{enumerate}

To the best of our knowledge, the best (fully covariant) decomposition method derives from the idea, 
first introduced by J. Kijowski \cite{K1}, of a ``canonical formalism'' where the metric is regarded as the 
``conjugate momentum'' to the connection in a purely affine Lagrangian theory 
and the second--order form of the equations is obtained through some kind of generalized 
Legendre transformation. The idea to generalize the Legendre transformation in this direction is 
mathematically correct and was applied to physics in the book \cite{KiTu} and later appeared many times in 
the literature. The choice of this formalism is strongly supported by a theorem that the dynamics of the 
unifying metric with the generic Lagrangian is equivalent to equations of motion for the 
resulting multiplet of fields and these equations necessarily contain the Einstein field 
equations for (the original or a transformed) metric \cite{K2}. 

In the usual Legendre transformation, the first step consists in the definition of conjugate momenta as function of 
the ``velocities'' (i.e.~the highest derivatives of the dynamical variables occurring in the Lagrangian). 
Then one should (at least locally) invert the mapping, so that 
all ``velocities'' in the Lagrangian can be replaced by suitable functions of the momenta. 

However, if one starts from a covariant second--order Lagrangian and aims at a covariant decomposition, 
the role of  ``velocities''cannot be played by the second derivatives of the metric components. In fact, the 
corresponding ``conjugate momenta'' would not be tensors; moreover, it would be impossible to invert the relation 
between ``velocities'' and ``momenta'', because a covariant Lagrangian can depend on second derivatives of a metric 
only through the components of the Riemann curvature. Then, one should consider a more general setting and introduce 
the conjugate momenta to the curvature components.To this end we first briefly recall the procedure introduced 
in \cite{MFF2,MFF3}. 

Consider a Lagrangian $L=L(\phi, \partial\phi, \partial^2\phi)$, 
where $\phi$ is a tensor field of some rank. $L$ is assumed to depend on a set of $k$ independent 
\textit{linear\/} combinations $\omega_A$, $A=1,2,\ldots,k$, of the second derivatives 
$\partial_{\alpha}\partial_{\beta}\phi$, i.e.~$L=L[\phi, \partial\phi, \omega_A(\phi, 
\partial\phi, \partial^2\phi)]$. Provided that $L$ is \textit{regular\/} in $\omega_A$, 
\begin{displaymath}
\det\left(\frac{\partial^2 L}{\partial\omega_A \partial\omega_B}\right)\neq 0
\end{displaymath}
(the Hessian matrix is of rank $k$), one defines momenta canonically conjugate to 
$\omega_A$,
\begin{displaymath}
\pi^A\equiv \frac{\partial L}{\partial\omega_A}
\end{displaymath}
and these equations can be inverted (solved w.r.t. $\omega_A$), $\omega_A=\Omega_A
(\phi, \partial\phi, \pi)$. Then one constructs an auxiliary function $H$ playing here a 
role analogous to the Hamiltonian in the standard canonical formalism of a field theory 
(this function should not be confused with the physical Hamiltonian generating the 
time evolution of the fields), 
\begin{displaymath}
H(\phi, \pi)\equiv \pi^A\Omega_A-L[\phi, \partial\phi, \Omega(\phi, \partial\phi, \pi)]
\end{displaymath}
and a Helmholtz Lagrangian (originally introduced by Helmholtz in mechanics \cite{He, Po, 
LCA})
\begin{displaymath}
L_H(\phi, \pi, \omega) \equiv \pi^A\omega_A-H(\phi, \pi).
\end{displaymath}
The dynamical variables are now $\phi$ and $\pi^A$ and $L_H$ is a specific Lagrangian depending 
on $\phi$, $\partial\phi$ and $\pi^A$, but not on $\partial\pi^A$. The variation $\delta\pi^A$ gives 
rise to second order equations 
\begin{displaymath}
\frac{\delta L_H}{\delta\pi^A}=0 \qquad \Rightarrow \qquad \omega_A=\Omega_A,
\end{displaymath}
whereas the fact that $L_H$ is linear in $\omega_A$ (and in consequence in $\partial^2\phi$) 
makes the equations $\delta L_H/\delta\phi=0$ third order ones. The two sets of equations 
for $L_H$ are equivalent to fourth order Lagrange equations for the original $L$, showing that 
this generalization of a Legendre transformation makes sense. Nevertheless it raises the problem: 
in the standard Hamiltonian formalism the canonical momenta are derivatives of a Lagrangian with respect 
to time derivatives of dynamical variables and carry independent degrees of freedom. The conjugate 
momenta defined above comprise both first and second time and spatial derivatives, then can they 
be carriers of physical DOF, or are they merely mathematically convenient auxiliary fields? \\
We claim (and this becomes evident when the theory of gravitational multiplets is developed 
in the following sections) that the equations arising from the Helmholtz Lagrangian involve 
derivatives of these auxiliary fields that cannot be integrated to give algebraic equations. 
Therefore we have all the reasons to treat these conjugate momenta (except $\chi$ below) as 
\textsl{bona fide} physical degrees of freedom.

The case of a generic NLG theory with $L=f(g_{\mu\nu}, R_{\alpha\beta\mu\nu})$ is more 
complicated than that above. The curvature tensor has altogether 14 independent 
scalar invariants and if one wishes to study NLG theories in full generality, then $f$ is 
a generic (transcendental) smooth function of these 14 variables. It is clear that if a 
theory is to be of any practical use, its Lagrangian should be 
simple. Long ago Einstein expressed this idea as a principle stating that Lagrangians 
of verified theories are the simplest possible functions of the field variables. In the 
case of NLG theories this principle is broken by definition, nevertheless their Lagrangians 
should be possibly simple and our task --- proving that the Weyl tensor should not be 
present in $L$ --- requires a proof for quite simple $f$, without entangling into a 
generic $L$. In the case of $L=f(R)$ it was shown \cite{MS1, S1, S2} using physical 
arguments that if $f(R)$ is analytic at $R=0$ (as it should be), then it should be of 
the form $R+aR^2+\ldots$, with $a>0$. For this Lagrangian it is not easy to find out constraints 
following from the field equations and in order to determine the number of DOF one first 
applies a version of this canonical formalism to decompose the unifying metric into a 
doublet comprising the metric and a scalar; this system has three DOF. This system is 
dynamically consistent showing that no DOF have been lost due to the appropriate 
generalized Legendre transformation. In our opinion the fact that it is quite easy to count the DOF 
for the doublet shows the advantage of dealing with these two fundamental fields instead of the 
unifying metric.

If $L$ explicitly depends on the Ricci tensor, one 
views it as an additive `correction' to $f(R)$: since any linear dependence on $R_{\mu\nu}$ is already 
incorporated in the term $f(R)$,  
the simplest additional term containing the Ricci tensor is quadratic; the quadratic term $R_{\mu\nu}R^{\mu\nu}$, 
on the other hand, should necessarily be present to ensure that the Lagrangian be regular with respect to $R_{\mu\nu}$ 
around the ground state (flat) solution.
The resulting Lagrangian $L=\kappa R+aR^2+bR_{\mu\nu}R^{\mu\nu}$ was studied in \cite{MS2}. It 
is clear that for this $L$ the linear combinations $\omega_A(\partial^2g_{\mu\nu})$ are 
$R$ and the traceless $S_{\mu\nu}\equiv R_{\mu\nu}-R g_{\mu\nu}/4$.

Next, if one admits a further correction due to the Weyl tensor, this should be expected to depend on the simplest Weyl 
invariant, $W\equiv \frac{1}{2}C_{\alpha\beta\mu\nu} C^{\alpha\beta\mu\nu}$; however, this invariant by itself does not 
provide a useful additional term, because (in dimension four) it can always be removed from the action integral by 
subtracting a full divergence (in accordance with the Gauss--Bonnet theorem). Even in higher dimensions, however,
 constraints arising from Bianchi identities are known to affect the DOF of a quadratic Lagrangian containing the Weyl 
tensor \cite{MFF3}). We shall therefore investigate $L$ of the form
\begin{equation}\label{n1}
L=\kappa R +aR^2 +bR_{\mu\nu}R^{\mu\nu}+f(W),
\end{equation}
where $f$ is a nonlinear function $(f''(W)\neq 0)$,
and we shall argue that the term $f(W)$ is physically redundant.   
We claim that this simple Lagrangian is sufficient for our task and that it is unlikely 
that a more intricate dependence of $L$ on $R$, $R_{\mu\nu}$ and $C_{\alpha\beta\mu\nu}$ 
will make the Weyl tensor dependence physical. In general the term, $f(W)$ may be dealt 
with in two different ways:\\
(i) either one defines a tensorial momentum conjugate to the linear combination of the second 
derivatives $\omega_A=C^{\alpha}{}_{\beta\mu\nu}$ (sections 5 and 7),\\
(ii) or one defines a scalar momentum conjugate to $W$ (sections 6 and 8).\\
We shall show that none of these ways provides a satisfactory physical theory (independently of whether 
some solutions may fit the observational data). 

The method referred as (ii) requires a preliminary explanation.  The generalized Legendre transformation, 
described above in this section, relies on the assumption that
the Lagrangian depends (in a regular way) on a \emph{linear} combination of the highest (in our case, the second) 
derivatives of a dynamical variable. To our knowledge, the definition of a momentum conjugate to a \emph{nonlinear} 
(e.g.~quadratic) function of the highest derivatives has not been considered so far. Yet, it works: the procedure yields 
an equivalent Lagrangian, but there is a substantial difference w.r.to a genuine Legendre transformation: here, the highest 
derivatives do not disappear, but the dependence of the Lagrangian on these derivatives is ``standardized''. To clarify 
the procedure, let us show a simple (first--order) example from classical particle mechanics.

On a finite--dimensional configuration space $Q$, consider a holonomic system described 
by a Lagrangian of the form
\begin{equation}\label{n2}
L(q^\lambda,\dot{q}^\lambda)=f(K)+U(q^\lambda)
\end{equation}
where $K\equiv\frac{1}{2}g_{\mu\nu}\dot{q}^\mu\dot{q}^\nu$ (the Lagrangian for an ordinary 
mechanical system corresponds to the case where $f$ is linear). For simplicity, assume that 
the configuration space metric $g_{\mu\nu}$ is constant. The Euler--Lagrange equations are
\begin{equation}\label{n3}
\frac{d}{dt}\left(f'(K)\cdot g_{\mu\nu}\dot{q}^\mu\right)-\frac{\partial U}{\partial q^\nu}=0.
\end{equation}
Let us now introduce the variable
\begin{equation}\label{n4}
p=\frac{\partial L}{\partial K}=f'(K).
\end{equation}
Provided $f''(K)\neq 0$, this relation can be inverted, i.e.~one can locally define a function 
$S(p)=K$ such that $f'(S(p))\equiv p$. Let us then introduce the Helmholtz Lagrangian
\begin{equation}\label{n5}
L_H(q^{\lambda},\dot{q}^{\lambda},p)=\frac{1}{2}p\, g_{\mu\nu}\dot{q}^{\mu}\dot{q}^{\nu} +V(p) + 
U(q^{\lambda}),
\end{equation}
where $V(p)=  f(S(p)) - p\,S(p)$. The Euler--Lagrange equations for $L_H$ are 
\begin{eqnarray}\label{n6}
\frac{1}{2}g_{\mu\nu}\dot{q}^{\mu}\dot{q}^{\nu}+\left(f'(S)-p\right)\frac{dS}{dp}-S(p) & = & 0 
\nonumber\\
\frac{d}{dt}\left(f'(S)\, g_{\mu\nu}\dot{q}^{\mu}\right)-\frac{\partial U}{\partial q^{\nu}} & = & 0
\end{eqnarray}
and are manifestly equivalent to (3), on account of $f'(S(p))= p$.

We have not added a new degree of freedom, because $L_H$ is degenerate: it does not depend on $\dot{p}$, 
and therefore the first of the E-L equations is actually a constraint. We have merely shown that the the 
original Lagrangian $L$, which depends in a nonlinear way on the kinetic energy $K$ (and thus is not 
quadratic in the velocity components $\dot{q}^\lambda$) is equivalent to a Lagrangian $L_H$ which \emph{is} 
quadratic in the velocity components, but is written in an extended configuration space $Q\times \mathbb{R}$: 
the number of degrees of freedom is unchanged because all solutions in $T(Q\times \mathbb{R})$ entirely lie 
on the submanifold defined by the equations $p=f'(K)$ and $\dot{p}=f''(K)\dot{K}$, which has the same 
dimension as $TQ$.

The observation above should be contrasted with the proper Legendre transformation. For a generic 
Lagrangian $L=L(q^{\lambda},\dot{q}^{\lambda})$, one sets $p_\mu=\partial L/\partial \dot{q}^{\mu}$,
and if the Lagrangian is hyperregular, i.e.~a global inverse Legendre map
\mbox{$\dot{q}^{\mu}=W^{\mu}(q^{\lambda},p_{\lambda})$} exists, one ends up with the Helmholtz Lagrangian
\begin{equation}\label{n7}
L_H(q^{\lambda},\dot{q}^{\lambda},p_{\lambda})=p_{\mu}\dot{q}^{\mu} - p_{\mu} W^{\mu}+
L(q^{\lambda},W^{\lambda}).
\end{equation}
The Helmholtz Lagrangian is now a function on $T(T^*Q)$. Apparently, the configuration space $Q$ has 
been enlarged to $T^*Q$, but the Lagrangian is now doubly degenerate: not only it does not depend on 
$\dot{p}_{\mu}$, but it is also \emph{linear} in $\dot{q}^{\mu}$. Therefore, the Euler--Lagrange 
equations are first--order (instead of second--order) in both $q^{\lambda}$ and $p_{\lambda}$. These 
equations define a vector field on $T^*Q$, while the E--L equations for the original Lagrangian $L$ 
(assumed to be hyperregular) define a vector field on $TQ$. 
In other words, after the Legendre transformation, the momenta $p_{\lambda}$ have completely replaced the velocity 
components $\dot{q}^{\lambda}$ as coordinates in the space of physical states of the system. \\
On the contrary, in the Helmholtz Lagrangian (5) the velocities $\dot{q}^{\lambda}$ are still present, but the new 
Lagrangian depends linearly on $K$: the original nonlinear dependence $f(K)$ in (2) has been ``absorbed'' by the 
auxiliary (non-dynamical) variable $p$; the coupling between $p$ and $K$ in (5) is ``universal'', and the only vestige 
of the original form of the Lagrangian is the ``potential'' $V(p)$ which generates a constraint.

In sections 6 and 8 we exploit this method to show that the scalar momentum conjugate to $W$ contributes to a 
mathematically correct dynamical description of the corresponding gravitational multiplet. It is a purely 
physical argument, rather than a mathematical one, that shows that the momentum conjugate to $W$ should be rejected.\\
In conclusion we state that the formalism developed in this work is applied to prove that if a metric 
nonlinear gravity theory such as that investigated here is not equivalent to GR plus a number of fields which 
represent non-geometric, physically distinct DOF, then the theory is not tenable from the field-theoretical 
viewpoint. In precisely this sense we shortly say that it is unphysical. And we emphasize that contrary to the 
common belief this equivalence does not hold for a generic gravitational Lagrangian.

\section{Lagrangian quadratically depending on $R$ and the Ricci tensor}
This case was discussed in detail in \cite{MS2} and here we present a modified version of those 
calculations focused on properties of the resulting gravitational triplet. There is no matter and 
the Lagrangian contains three coupling constants, 
\begin{eqnarray}\label{n8}
&
L=\kappa R+aR^2+bR_{\mu\nu}R^{\mu\nu}=
& \nonumber\\
&
=\kappa R+\left(a+\dfrac{b}{4}\right)R^2+bS_{\mu\nu}S^{\mu\nu}.
&
\end{eqnarray}
Here
\begin{equation}\label{n9}
R_{\mu\nu}=S_{\mu\nu}+\frac{1}{4}g_{\mu\nu}R, \quad \kappa\equiv\frac{c^3}{16\pi G}.
\end{equation}
Positivity of energy in Einstein frame for $L=f(R)$ gravity requires $a>0$ \cite{MS1}, 
whereas it is suggested in \cite{MS2} that one should have $b<0$ and $3a+b\geq 0$, for the reasons that we 
recall below. Dimensionalities are: 
$[L]=[\kappa R]=M L^{-2} T^{-1}$, $[\kappa]=M T^{-1}$, $[a]=[b]=M L^2 T^{-1}$ and $[\kappa/b]=L^{-2}$. The 
fourth-order field equations were first (?) derived in \cite{KS} and then in many other works (see e.g. 
\cite{MS2, LP}); clearly they are equivalent to second-order equations derived below. 
One introduces one scalar and one tensor momentum, 
\begin{equation}\label{n10}
\chi\equiv \frac{\partial L}{\partial R}-\kappa=2\left(a+\frac{b}{4}\right)R,
\end{equation}
\begin{equation}\label{n11}
\pi^{\mu\nu}\equiv \frac{\partial L}{\partial S_{\mu\nu}}=2b S^{\mu\nu},
\end{equation}
hence $\pi^{\mu\nu}g_{\mu\nu}=0$. The fields $\chi$ and $\pi^{\mu\nu}$ are of the same dimensionality 
equal to $[\chi]=[\pi^{\mu\nu}]=[\kappa]=M T^{-1}$. For convenience we set $c=1$. In this way the 
original metric gets decomposed into a triplet forming the Helmholtz--Jordan frame (HJF) 
$\{g_{\mu\nu}, \pi^{\mu\nu},\chi\}$. The metric remains unchanged (the signature is $-+++$), only 
its dynamics will be described by equations of motion of different form. 
The function $H$ (pseudo--Hamiltonian) is 
\begin{eqnarray}\label{n12}
&
H \equiv \frac{\partial L}{\partial R}(\chi) R(\chi) +
\frac{\partial L}{\partial S_{\alpha\beta}}(\pi) S_{\alpha\beta}(\pi)- L(g, \chi, \pi)=
& \nonumber\\
&
= \frac{1}{4a+b}\chi^2 +\frac{1}{4b}\pi^{\alpha\beta} \pi_{\alpha\beta}. 
&
\end{eqnarray}
It generates the Helmholtz Lagrangian
\begin{eqnarray}\label{n13}
&
L_H(g, R, \chi, S_{\mu\nu}, \pi) \equiv  
\frac{\partial L}{\partial R}(\chi) R(g) +
\frac{\partial L}{\partial S_{\alpha\beta}}(\pi) S_{\alpha\beta}(g)-H=
& \nonumber\\
&
=\kappa R + \chi R + \pi^{\alpha\beta}S_{\alpha\beta} -
\frac{1}{4a+b}\chi^2 -\frac{1}{4b}\pi^{\alpha\beta} \pi_{\alpha\beta}.
&
\end{eqnarray}
The Lagrange equations for the momenta $\chi$ and $\pi^{\mu\nu}$ are algebraic 
and recover the definitions of the fields, 
\begin{equation}\label{n14}
\frac{\delta L_H}{\delta\chi}=0 \Rightarrow R=\frac{2}{4a+b}\chi,
\end{equation}
\begin{equation}\label{n15}
\frac{\delta L_H}{\delta\pi^{\mu\nu}}=0 \Rightarrow S_{\mu\nu}=\frac{1}{2b}\pi_{\mu\nu},
\end{equation}
eq. (15) confirms that $\pi_{\mu\nu}$ is traceless. Equations (14) and (15) are together composed into 
quasi--Einsteinian field equations, 
\begin{equation}\label{16}
G_{\mu\nu}=\frac{1}{2b}\pi_{\mu\nu}-\frac{1}{2(4a+b)}\chi g_{\mu\nu},
\end{equation}
and these give rise to four differential constraints (due to Bianchi identities) 
\begin{equation}\label{n17}
\pi_{\mu\nu}{}^{;\nu}-\frac{b}{4a+b} \chi_{;\mu}=0.
\end{equation}
The metric variation of $L_H$ generates EFE $G_{\mu\nu}=8\pi G T_{\mu\nu}(g, \chi, \pi)$ comprising an 
energy--momentum tensor for the two momenta. One sees comparing (16) and EFE that for solutions 
these equations provide two expressions for $T_{\mu\nu}$ which may be used both as equations for the 
metric and for the momenta $\chi$ and $\pi_{\mu\nu}$. Dividing (16) by $8\pi G$  one gets
\begin{eqnarray}\label{n18}
T_{\mu\nu} & = & \frac{1}{16\pi Gb}(\pi_{\mu\nu}-\frac{b}{4a+b}\chi g_{\mu\nu})=
\nonumber\\
& = & 2\chi_{;\mu\nu}+2\pi^{\alpha}{}_{(\mu;\nu);\alpha}-\pi_{\mu\nu;\alpha}{}^{;\alpha}
+g_{\mu\nu}[-2\Box\chi-\pi^{\alpha\beta}{}_{;\alpha\beta}+\frac{1}{4b}\pi^{\alpha\beta}
\pi_{\alpha\beta}]-
\nonumber\\
& - & \frac{1}{b}\pi^{\alpha}{}_{\mu}\pi_{\alpha\nu}-\frac{2(2a+b)}{b(4a+b)}\chi \pi_{\mu\nu}.
\end{eqnarray}
First the trace of (18) generates an equation of motion for $\chi$. Applying the divergence to eq. (17) and  
inserting the resulting expression into this equation one gets 
\begin{equation}\label{19}
(3a+b)\Box\chi-\frac{1}{32\pi G}\chi=0.
\end{equation}
For $3a+b>0$ the mass of $\chi$ is real, $m^2_{\chi}=[32\pi G(3a+b)]^{-1}$. \\
If $3a+b=0$ the scalar $\chi$ is constrained to vanish \cite{MS2}. Now assuming $3a+b>0$ for $a>0$ and $b<0$ 
(negative $b$ gives rise to real mass for $\pi_{\mu\nu}$ in the case $3a+b=0$ \cite{MS2}) one can 
eliminate $\chi$ from (18). To this end one again takes the divergence of (17) which reads
\begin{equation}\label{20}
\Box\chi=\frac{4a+b}{b} \pi^{\alpha\beta}{}_{;\alpha\beta}
\end{equation}
and applies it to eliminate $\Box\chi$ from (19) and finally arrives at  
\begin{equation}\label{21}
\chi=\frac{32\pi G}{b}(3a+b)(4a+b) \pi^{\alpha\beta}{}_{;\alpha\beta}.
\end{equation}
This shows that $\chi$ is not an independent physical field, being completely determined by values of 
the double divergence of $\pi^{\mu\nu}$. One would be lead to regard the scalar momentum as being spurious 
and remove it by the appropriate restriction of Lagrangian (8), $3a+b=0$. However, (\ref{21}) was obtained 
using (\ref{16}), so this relationship between $\chi$ and $\pi^{\mu\nu}$ might be an outcome of their 
interaction with the metric. As we have discussed in the previous 
section, to prove that this is not the case we need to envisage the behavior of these fields in 
the decoupling limit in flat spacetime.

Before taking this limit we exhibit the complete dynamical structure of the model. To this end one replaces 
in eqs. (18) $\chi$ by (21), $\chi_{;\mu\nu}$ by the derivative of (17) and $\Box\chi$ by (20) and arrives at 
\begin{eqnarray}\label{n22}
T_{\mu\nu} & = & \frac{1}{16\pi Gb}\pi_{\mu\nu}-\frac{2}{b}(3a+b)
\pi^{\alpha\beta}{}_{;\alpha\beta} g_{\mu\nu}=
\nonumber\\
& = & \frac{2}{b}(4a+b)\pi_{\alpha(\mu}{}^{;\alpha}{}_{;\nu)}+
2\pi^{\alpha}{}_{(\mu;\nu);\alpha}-\pi_{\mu\nu;\alpha}{}^{;\alpha}-
\frac{1}{b}\pi^{\alpha}{}_{\mu}\pi_{\alpha\nu}+
\nonumber\\
& + & g_{\mu\nu}[-\frac{1}{b}(8a+3b)\pi^{\alpha\beta}{}_{;\alpha\beta}+
\frac{1}{4b}\pi^{\alpha\beta}\pi_{\alpha\beta}]-
\nonumber\\
& - & 
\frac{64\pi G}{b^2}(2a+b)(3a+b)\pi^{\alpha\beta}{}_{;\alpha\beta}\pi_{\mu\nu}.
\end{eqnarray} 
Notice that the symbol $T_{\mu\nu}$ is shown here merely to indicate the origin of this equation, but is 
otherwise irrelevant. After elimination of the scalar the system of equations of motion consists of:\\
--- 9 quasi--linear eqs. (22) for $\pi_{\mu\nu}$ (their trace vanishes identically),\\
--- EFE eqs. (16) for $g_{\mu\nu}$,\\
--- expression (21) for the auxiliary scalar, to be inserted into (16).\\
These equations do not directly follow from a Lagrangian according to the standard formalism, instead one 
has to perform the above more involved procedure. The fact that the procedure generates two different 
expressions for $T_{\mu\nu}(g,\chi,\pi)$ which are equal for solutions, plays essential role in it. These 
equations are independent. In fact, (16) successively generate (17) and (20), but one cannot derive 
(21) from them and from (22). Yet applying (21) to (16) and (22) one can invert the whole procedure. The 
fact that equation (19) can be integrated to yield (21) shows that the initial data for $\chi$ are 
determined by initial data for $\pi_{\mu\nu}$ and the scalar carries no its own degrees of freedom. 
It is likely that the possibility of integrating of (19) is due 
to the fact that the Lagrangian is quadratic in $R$ and $R_{\mu\nu}$; in presence of higher order terms 
an analogous equation for $\chi$ is not expected to be integrable to an algebraic expression. Yet it is 
unclear at this point whether an algebraic relationship such as (21) does imply that the number of 
available DOF for the doublet $\{g_{\mu\nu}, \pi_{\mu\nu}\}$ is decreased from 8 to 7. This issue must 
be determined in the decoupling limit in flat spacetime and we shall see that there are still 8 DOF 
(including the metric).\\

To get the physical interpretation we now consider the \textit{gravity--free\/} system in flat 
spacetime. Since the system of equations (16), (22) and (21) is equivalent to (16), (17), (18) and (20), 
we take the decoupling limit in the first system and parallelly observe its outcomes in the latter system.\\
1. By analogy to taking the limit in GR coupled to ordinary matter, we first decouple all physical fields 
from the spacetime metric by putting $G\rightarrow 0$ in EFE $G_{\mu\nu}=8\pi G T_{\mu\nu}$. Here 
$T_{\mu\nu}$ is the variational energy-momentum tensor generated by $L_H$; in principle it might include 
the contribution from ordinary matter. In this limit the spacetime is a fixed solution to   
$G_{\mu\nu}=0$ and $T_{\mu\nu}$ describes test fields in it.\\
2. Vanishing of $G_{\mu\nu}$ implies that RHS of (16) vanishes for arbitrary 
$\pi_{\mu\nu}$ and $\chi(\pi)$. This is possible only in the limit $1/b\rightarrow 0 \Leftrightarrow b\rightarrow 
-\infty$. Formally it is sufficient and the constant $a$ might remain finite, $|a|<\infty$, but it 
would break the condition $3a+b>0$. This shows that also $a$ must go to infinity and at the same 
rate as $b$. Let $a=-\xi b$ for dimensionless $\xi$, then $\xi>1/3$ and let $\xi$ remain constant 
for $b\rightarrow -\infty$. In this limit  eqs. (14) and (15) reduce to $R(g)=0$ and 
$S_{\mu\nu}(g)=0$. \\
3. To take the limit of $G$, $a$ and $b$ in formula (21) (and respectively in (19)) one assumes that $G$ 
and $1/b$ tend 
to 0 at the same rate, so that the product $Gb$ is finite and negative in this limit. Let 
$\kappa/b\rightarrow -\lambda^{-2}<0$ with $[\lambda]=L$. One finds in the limit that eq. (19) reads
\begin{equation}\label{23}
\Box\chi=\frac{\chi}{2(3\xi-1)\lambda^2}
\end{equation}
and formula (21) takes the form
\begin{equation}\label{24}
\chi=-2(3\xi-1)(4\xi-1)\lambda^2 \pi^{\alpha\beta}{}_{;\alpha\beta},
\end{equation}
showing that also in the gravity-free system the scalar is not independent and is just a name for the 
double divergence of $\pi_{\mu\nu}$. For further use we show the constraint (17) which reads now 
\begin{equation}\label{n25}
\pi_{\mu\nu}{}^{;\nu}=\frac{1}{1-4\xi}\chi_{;\mu}.
\end{equation}
4. Finally one takes the limit of the three coupling constants in eq. (22). After some manipulations 
one arrives at 
\begin{equation}\label{26}
E_{\mu\nu}\equiv -4(1-2\xi)\pi_{\alpha(\mu}{}^{;\alpha}{}_{;\nu)}+\pi_{\mu\nu;\alpha}{}^{;\alpha}-
(2\xi-1)g_{\mu\nu}\pi^{\alpha\beta}{}_{;\alpha\beta}-\frac{1}{\lambda^2}\pi_{\mu\nu}=0.
\end{equation}
The whole system gets reduced to a test linear massive field $\pi_{\mu\nu}$ subject to eqs. (26), whereas 
one preserves eqs. (23), (24) and (25) as auxiliary formulae to be applied later on.\\

Now one performs the second procedure of the two fundamental ones described in Sect.2, namely one studies 
the dynamics in flat spacetime. In this spacetime one may determine the physical properties of 
$\pi^{\mu\nu}$ by solving the above system of equations in 
terms of arbitrary initial data. In a chosen inertial reference frame and using Cartesian coordinates 
one gives initial data at $t=0$: $\pi_{\mu\nu}(t=0)=f_{\mu\nu}(\mathbf{x})$ and $\partial_0\pi_{\mu\nu}
(t=0)=h_{\mu\nu}(\mathbf{x})$. These are together 18 functions since $\pi_{\mu\nu}\eta^{\mu\nu}=0=
\pi_{\mu\nu,0}\eta^{\mu\nu}$ imply $f_{00}=\sum_j f_{jj}$ and $h_{00}=\sum_j h_{jj}$ for $i,j=1,2,3$. 
To show that these Cauchy data uniquely determine a solution to eqs. (26) one first 
solves eq. (23) for $\chi$ as if it were an independent variable, applying (25). The 
initial data define a scalar and a 3--vector at $t=0$, 
\begin{equation}\label{27}
\pi_{0\nu}{}^{,\nu}=\sum_j(-h_{jj}+f_{0j,j})\equiv H(\mathbf{x}) \quad \textrm{and} \quad 
\pi_{i\nu}{}^{,\nu}=-h_{0i}+\sum_j f_{ij,j}\equiv K_i(\mathbf{x}),
\end{equation}
these are known functions. The constraints (25) determine initial data for $\chi$, 
\begin{equation}\label{28}
\chi_{,0}=(1-4\xi)H(\mathbf{x}), \qquad \partial_i\chi=(1-4\xi)K_i(\mathbf{x}).
\end{equation}
One sees that $K_i$ must be a gradient and the functions $f_{ij}$ and $h_{0i}$ are subject to 
$K_{i,j}=K_{j,i}$. If these constraints hold, the components $K_i$ uniquely determine (up to an 
additive constant) the function $K(\mathbf{x})$ such that $K_i=\partial_i K$. Then the initial value 
of $\chi$ is $\chi(t=0)=(1-4\xi)K(\mathbf{x})$ and the integration constant is eliminated by imposing 
appropriate boundary conditions at infinity on $h_{0i}$, $f_{ij}$ and $\chi(0)$. The Cauchy data  
$\chi(t=0)$ and $\chi_{,0}$ uniquely determine the solution $\chi(x^{\mu})$ of (23). 
Now one returns to eqs. (26) and employing (24) and (25) one finds that in flat space they read
\begin{eqnarray}\label{n29}
E_{\mu\nu} & = & \pi_{\mu\nu,\alpha}{}^{,\alpha}-\frac{1}{\lambda^2}\pi_{\mu\nu}-
\frac{4(1-2\xi)}{1-4\xi}\chi_{,\mu\nu}+\frac{1-2\xi}{2(1-4\xi)(3\xi-1)\lambda^2} \eta_{\mu\nu}
\chi \equiv
\nonumber\\
& \equiv & \pi_{\mu\nu,\alpha}{}^{,\alpha}-\frac{1}{\lambda^2}\pi_{\mu\nu}+\tau_{\mu\nu}(\chi)=0.
\end{eqnarray}
Although $\chi$ fully depends on $\pi_{\mu\nu}$, we have seen that the solution for $\chi$ can be fully 
determined from initial data, without knowing the solution for $\pi_{\mu\nu}$: doing so, the term $\tau_{\mu\nu}$ 
becomes an \textit{explicitly known\/} source for unknown $\pi_{\mu\nu}$. Physically this is a bizarre situation, 
yet mathematically there is nothing inconsistent here; (29) is a kind of tensorial Klein--Gordon equation and 
its solution is uniquely determined by the initial data.

Having this knowledge one counts the degrees of freedom for $\pi_{\mu\nu}$. The constraints 
$K_{i,j}=K_{j,i}$ reduce the number of independent functions $K_i$ to one, $K(\mathbf{x})$, hence the number of 
independent data is diminished by 2. Then there are dynamical constraints since not all of $E_{\mu\nu}
=0$ are hyperbolic propagation equations. From (25) one finds that 
\begin{displaymath}
\pi_{00,0}=\sum_j \pi_{0j,j}-\frac{\chi_{,0}}{1-4\xi}, \quad \pi_{i0,0}=\sum_j \pi_{ij,j}-
\frac{\chi_{,i}}{1-4\xi}
\end{displaymath}
and four equations $E_{\mu0}=0$ read
\begin{equation}\label{n30}
\sum_j(\pi_{\mu0,jj}-\pi_{\mu j,0j})-\frac{1}{\lambda^2}\pi_{\mu0}+\tau_{\mu0}-\frac{1}{4\xi-1} 
\chi_{,\mu0}=0,
\end{equation}
hence these equations are constraints ones. However $E_{00}=0$ is not an independent equation, since 
$\eta^{\mu\nu} E_{\mu\nu}\equiv 0$ yields $E_{00}=\sum_j E_{jj}$ and the independent dynamical constraints 
are $E_{i0}=0$ and $\sum_j E_{jj}=0$. Together the number of independent initial data is $18-2-4=12$ 
corresponding to six DOF for $\pi_{\mu\nu}$. Yet a massive quantum spin--two particle has \textit{five\/} 
DOF \cite{FP, H, SH, T, ADY} indicating that the hypothetical particles of the quantized field 
$\pi_{\mu\nu}$ do not have a definite spin. This is unphysical.

In the case $3a+b=0$ the scalar momentum is eliminated by its equation of motion and $\pi^{\mu\nu}$ is 
subject to the tensorial Klein--Gordon equations with the mass $m_{\pi}=1/\lambda$ and constraints 
$\eta^{\mu\nu} \pi_{\mu\nu}=0=\partial_{\nu}\pi^{\mu\nu}$ in flat space. This field carries 5 DOF 
\cite{ADY} and the doublet $\{g_{\mu\nu}, \pi^{\mu\nu}\}$ has 7 DOF (since for $\pi^{\mu\nu}=0$ the metric 
satisfies $G_{\mu\nu}=0$ and carries two DOF). This agrees with the fact that the unifying metric has 
seven DOF: the original fourth order equations, beyond the standard constraints giving rise to 8 DOF, 
satisfy in this case two additional constraints, $R=0$ and $\partial R/\partial t=0$, which 
diminish the number to seven. For the Lagrangian (8) we previously assumed in \cite{MS2}, just for 
convenience, that $3a+b=0$, then we got dynamically $\chi=0$ and 7 DOF for the doublet. However, the 
choice  $3a+b=0$ is not a matter of convenience, we emphasize that it is required by physics: though 
the scalar $\chi$ inevitably appears at an intermediate step in the Legendre transformation formalism, 
it is not only spurious as being determined by $\pi^{\mu\nu}$, moreover it must vanish (due to equations 
of motion), otherwise one cannot assign a definite spin to $\pi^{\mu\nu}$.\\

For the sake of completeness we make a comment on physical properties of $\pi^{\mu\nu}$ in this case. Eq. (16)
shows that the field is not fully satisfactory physically: for solutions its energy--momentum tensor is linear, 
hence its energy density is indefinite. Its properties were investigated in \cite{MS2}. Its ground state 
solution $\pi^{\mu\nu}=0$ is at least linearly stable (the mass is positive if $b>0$); the problem of nonlinear 
stability (the exact nonlinear $\pi^{\mu\nu}$ interacts with the metric) is hard and open. Whether or not 
$\pi^{\mu\nu}$ is a ghost field cannot be established in the exact 
theory since its (Helmholtz) Lagrangian (13) does not contain at all derivative terms. The field turns out a 
ghost in a linear approximation \cite{MS2} and this is proven in a rather intricate way. Thus $\pi^{\mu\nu}$ is 
not a typical ghost. Furthermore there are arguments \cite{HH} that being a ghost is not so disastrous as it 
was formerly believed.

Conclusion: \textit{the gravitational theory based on (8) is equivalent to GR interacting with a massive 
spin-2 field only in the case $3a+b=0$ and only in this case may be tenable from the field--theoretical 
viewpoint\/}.

\section{Tensorial momentum conjugate to the Weyl tensor}

From the fact that both $L=f(g_{\mu\nu}, R_{\alpha\beta\mu\nu})$ and 
$L=R+aR^2+bR_{\mu\nu}R^{\mu\nu}$ (prior to excluding the scalar) have 8 DOF (in the 
mathematical sense) one would be tempted to conclude that the presence of the Weyl tensor is 
redundant. Yet, we have just shown that the latter Lagrangian cannot represent a physical field theory with 8 DOF, so 
one may wonder if adding to (8) (with $3a+b=0$) a term involving the conformal tensor would restore the expected 8 DOF. 

We thus apply the approach developed in previous sections to a Lagrangian of the form (1), with $3a+b=0$: 
\begin{equation}\label{n31}
L=\kappa R +\frac{1}{3m^2}(R^2 -3R_{\mu\nu}R^{\mu\nu})+\frac{1}{k} f(W);
\end{equation}
the theory has three coupling constants. Here $[m^2]= M^{-1}L^{-2} T$ and 
$W\equiv \frac{1}{2}C_{\alpha\beta\mu\nu} C^{\alpha\beta\mu\nu}$ 
with $[W]=L^{-4}$; $f$ is a smooth function. The dimension of $k$ depends 
on the form (dimensionality) of $f$ so that $[\frac{1}{k}f(W)]$ is the same as $[\kappa R]$. \\
We assume that $f(W)$ contains no dimensional constants and that $f(W)$ is analytic at $W=0$, 
$f(0)=0=f'(0)$ and $f''(W)\neq 0$.

In the sequel, to avoid making the computations pointlessly cumbersome we assume that $f(W)$ is a simple function, say 
\begin{equation}\label{n32}
f(W)=\frac{1}{n}\,W^n, 
\end{equation}
with $n$ some even positive integer: this allows us to invert $f'(W)$ explicitly and thus to spell out all terms 
in the subsequent computations. The theory for the momentum $\sigma_{\alpha}{}^{\beta\mu\nu}$ should be consistent for 
any such $n$.

The ``traditional'' strategy that we have adopted so far consisted in introducing conjugate momenta which carry definite spin. 
This suggests to introduce again the scalar and the tensor momenta conjugate to the Ricci tensor, $\chi$ and $\pi^{\mu\nu}$ 
respectively, as in (10) and (11) (the unwanted scalar momentum may be eliminated only at the level of equations of motion), 
and in addition a new tensorial conjugate momentum corresponding to the Weyl tensor \cite{MFF3}: 
 \begin{equation}\label{n33}
\sigma_{\alpha}{}^{\beta\mu\nu}\equiv \frac{\partial L}
{\partial C^{\alpha}{}_{\beta\mu\nu}} =\frac{\partial L}{\partial W}
\frac{\partial W}{\partial C^{\alpha}{}_{\beta\mu\nu}} =\frac{1}{k} f'(W) C_{\alpha}{}^{\beta\mu\nu}.
\end{equation}
All momenta have dimensionality as $MT^{-1}$. Following from their definitions, $\pi^{\mu\nu}$ 
is traceless and $\sigma_{\alpha}{}^{\beta\mu\nu}$ has all symmetries of the conformal tensor: thus, at first sight, 
$\sigma_{\alpha}{}^{\beta\mu\nu}$ would be expected to correspond to a spin--two field as well. Yet, we know that 
the model has at most 8 DOF, so it is evident that there is no place for two independent spin--two fields; however, 
we need to understand how $\pi^{\mu\nu}$ and $\sigma_{\alpha}{}^{\beta\mu\nu}$ are constrained to each other (in the 
coupling-free limit where they should behave as test fields, if they were physically independent) to assess 
whether non--independent components can be identified and eliminated, to obtain -- if possible -- a viable covariant 
field theory which may saturate the 8 mathematically available DOF.

To invert the relationship (33) one introduces the square of $\sigma$, 
$Z \equiv \frac{1}{2}\sigma^{\alpha\beta\mu\nu}\sigma_{\alpha\beta\mu\nu}$ and finds that it equals 
\begin{equation}\label{n34}
Z=\frac{W}{k^2}[f'(W)]^2.
\end{equation}
We denote the inverse of the function (34) by $W=v(Z)$, hence
\begin{equation}\label{n35}
C^{\alpha\beta\mu\nu}(\sigma) = \frac{k}{f'(v(Z))} \sigma^{\alpha\beta\mu\nu}. 
\end{equation}
The function $H$ is 
\begin{eqnarray}\label{n36}
H(\chi, \pi, \sigma) & \equiv & \frac{\partial L}{\partial R}(\chi)R(\chi)+\frac{\partial L}{\partial 
S_{\mu\nu}}(\pi) S_{\mu\nu}(\pi)+\frac{\partial L}{\partial C^{\alpha}{}_{\beta\mu\nu}}(\sigma)
C^{\alpha}{}_{\beta\mu\nu}(\sigma)-
\nonumber\\
& - & L(g,\chi,\pi,\sigma) =
3m^2\chi^2 -\frac{m^2}{4}\pi^{\alpha\beta} \pi_{\alpha\beta}+\frac{2kZ}{f'(v(Z))} -\frac{1}{k} f(v(Z))
\end{eqnarray}
and the Helmholtz Lagrangian reads 
\begin{eqnarray}\label{n37}
L_H  & \equiv & \frac{\partial L}{\partial R}(\chi)R(g)+\frac{\partial L}{\partial S_{\mu\nu}}(\pi) 
S_{\mu\nu}(g)+\frac{\partial L}{\partial C^{\alpha}{}_{\beta\mu\nu}}(\sigma)
C^{\alpha}{}_{\beta\mu\nu}(g)-H(\chi, \pi, \sigma, \psi)=
\nonumber\\
& = & \kappa R + \chi R + \pi^{\alpha\beta}S_{\alpha\beta} + \sigma_{\alpha}{}^{\beta\mu\nu} 
C^{\alpha}{}_{\beta\mu\nu}-3m^2\chi^2 +\frac{m^2}{4}\pi^{\alpha\beta} \pi_{\alpha\beta} -\frac{2kZ}{f'(v(Z))} +
\nonumber\\
& + & \frac{1}{k} f(v(Z)).
\end{eqnarray}
$L_H$ does not contain derivatives of the momenta, hence the Lagrange eqs. for them are algebraic and recover 
their definitions,
\begin{equation}\label{n38}
R(g)=6m^2 \chi, \qquad S^{\mu\nu}(g)=-\frac{m^2}{2}\pi^{\mu\nu}
\end{equation}
and (35). Eqs. (38) combine together in quasi--Einsteinian field equations,
\begin{equation}\label{n39}
G_{\mu\nu}=-\frac{m^2}{2}(\pi_{\mu\nu}+3\chi g_{\mu\nu})
\end{equation}
and Bianchi identities imply 4 differential constraints 
\begin{equation}\label{n40}
\pi_{\mu\nu}{}^{;\nu}+3\chi_{;\mu}=0.
\end{equation}
While taking variations w.r.t. metric $g_{\mu\nu}$ one assumes, as always, that the 
components $\sigma_{\alpha}{}^{\beta\mu\nu}$, as they are defined in (33), are independent of 
the metric, $\delta_g \sigma_{\alpha}{}^{\beta\mu\nu}\equiv 0$. After a number of manipulations 
one arrives at the following system of equations of motion, again having the form of EFE. If we assume (32), these read
\begin{eqnarray}\label{n41}
\kappa G_{\mu\nu} -\frac{2}{3}\pi_{\alpha(\mu}{}^{;\alpha}{}_{;\nu)}+
\frac{1}{6}\pi^{\alpha\beta}{}_{;\alpha\beta}g_{\mu\nu}- 
\frac{m^2}{2}\chi \pi_{\mu\nu}+\frac{1}{2}\pi_{\mu\nu;\alpha}{}^{;\alpha}
-R_{\alpha\mu\nu\beta}\pi^{\alpha\beta} 
+\frac{m^2}{8}
\pi^{\alpha\beta}\pi_{\alpha\beta} g_{\mu\nu}-& &\nonumber \\ 
-2\sigma_{\alpha\mu\nu\beta}{}^{;\alpha\beta}+
\frac{m^2}{2}\sigma_{\alpha\mu\nu\beta}\pi^{\alpha\beta}+\frac{n-1}{2n}(kZ^n)^{\frac{1}{2n-1}} g_{\mu\nu}\equiv 
 \kappa G_{\mu\nu} -\frac{1}{2}T_{\mu\nu}(g,\chi,\pi,\sigma)=0.\qquad& &
\end{eqnarray}
All the terms in (41) explicitly depending on the three momenta should be interpreted as 
a collective energy--momentum tensor for them. In practice RHS of (39) is an effective total  
energy--momentum tensor $8\pi GT_{\mu\nu}$ expressed in terms of solutions.

The trace of equations (41) determines the scalar $\chi$ as a function of the scalar $Z(\sigma)$,
\begin{equation}\label{n42}
\chi =\frac{16\pi G}{3nm^2}(n-1)(kZ^n)^{\frac{1}{2n-1}},
\end{equation}
hence the scalar momentum is eliminated from the field equations (if $f(W)=0$, $\chi\equiv 0$). 
We emphasize that this elimination is possible due to the presence of two expressions for the 
energy--momentum tensor $T_{\mu\nu}$, which are equal for solutions. One can therefore employ (41) as 
propagation equations for the remaining two momenta, whereas eqs. (39) take the role of Einstein 
equations for the metric. (Whether or not this stress tensor has positive definite energy density 
may be determined if the space of solutions is known.) In the propagation equations one should 
replace $G_{\mu\nu}$ according to (39).

However eqs. (35) form another set of equations for the metric where the momentum $\sigma$ is a 
source for the Weyl tensor $C_{\alpha\beta\mu\nu}(g)$. Hence the full Riemann tensor is expressed 
as an algebraic function of the momenta. This shows that the formulated here second-order dynamics 
is different from the standard GR dynamics comprising EFE and equations of motion for interacting 
matter fields. Equations (35) are consistent with (39) provided $C_{\alpha\beta\mu\nu}$ 
satisfies the standard differential identities for the conformal tensor. The first order identity 
in four dimensions, upon employing (35), (38) and (32) reads 
\begin{eqnarray}\label{n43}
\lefteqn{
\frac{1-n}{2n-1}Z^{-1} Z^{;\nu}\sigma_{\alpha\beta\mu\nu}+\sigma_{\alpha\beta\mu\nu}{}^{;\nu}=}
\nonumber\\
& &
=\frac{m^2}{4}(k^{-1}Z^{n-1})^{\frac{1}{2n-1}}\left(\pi_{\mu\beta;\alpha}-\pi_{\mu\alpha;\beta}
+g_{\alpha\mu} \chi_{;\beta}-g_{\beta\mu} \chi_{;\alpha}\right).
\end{eqnarray}
Since the scalar $\chi$ is merely a function of the other fields, the degrees of freedom are associated 
with the gravitational triplet of fields $\{g_{\mu\nu},\pi^{\mu\nu}, \sigma_{\alpha\mu\nu\beta}\}$ and 
are subject to the following equations of motion:\\
--- 10 EFE (39), equivalent to (38),\\
--- 10 eqs. (35) which are not viewed as propagation equations for the metric and instead are 
interpreted as generating equations for necessary constraints imposed on the momenta,\\
--- the constraints (43),\\
--- 9 differential propagation eqs. (41) for 9 momenta $\pi^{\mu\nu}$ and 10 momenta 
$\sigma_{\alpha\mu\nu\beta}$ (one equation, the trace, has been used up to provide an algebraic 
expression eliminating the scalar $\chi$),\\
--- expression (42) for $\chi$.\\
Notice that for these equations there is no limit (such as $n\rightarrow 0$) which would reduce this 
system to that studied in sect. 4.\\
This is a determined system of equations of motion which are equivalent to the 10 fourth order equations 
for the metric $g_{\mu\nu}$ directly following from the Lagrangian (31).

Mathematically this is OK, yet this is inconsistent with field theory in flat spacetime 
requiring that each physical field, when its couplings to other fields are switched off, 
should be determined by a propagation equation (possibly containing self--interaction terms). In fact, 
to elucidate the issue, one decouples the two momenta from the metric, then it turns out that the 
decoupling makes the spacetime flat. As we shall see below it follows that $9+10$ fields $\pi^{\mu\nu}$ 
and $\sigma_{\alpha\mu\nu\beta}$ are underdetermined.
\begin{enumerate}[noitemsep]
\item First, one decouples all the momenta from the curvature by putting  
$G\rightarrow 0$ in eqs. (41) written in the form $G_{\mu\nu}-8\pi GT_{\mu\nu}=0$ 
and gets $G_{\mu\nu}=0$.\\
\item Consistency of the effective Einsteinian eqs. (39) with $G_{\mu\nu}=0$ requires taking 
the limit $m^2 \rightarrow 0$.\\
\item The momentum $\sigma_{\alpha\mu\nu\beta}$ is fully decoupled from the curvature if 
$k\rightarrow 0$ in eq. (35). Then $C_{\alpha\mu\nu\beta}=0$ and the spacetime is flat, 
$R_{\alpha\mu\nu\beta}=0$. In general relativity, if matter does not gravitate, $R_{\mu\nu}
=0$ and gravitational waves may exist. Here one sees that inclusion of the conformal tensor 
in the Lagrangian as an independent field in gravitational multiplet has very restrictive 
consequences.\\
\item The power of $k$ in (42) is positive and in the limit $k=0$ the scalar vanishes, $\chi=0$. 
In consequence the constraint (40) reads $\pi_{\mu\nu}{}^{;\nu}=0$.\\
\item To proceed further one assumes that $G$, $m^2$ and $k$ vanish at the 
same rate, so that both $m^2\kappa=m^2/(16\pi G)\equiv\lambda^{-2}$ and $m^2/k\equiv\mu$ 
remain constant and positive in this limit; $\lambda$ has the dimension of length. \\ 
\item Under this assumption the constant factor in RHS of (43) vanishes in the limit,
\begin{displaymath}
m^2\,k^{-\frac{1}{2n-1}}=\mu\,k^{\frac{2(n-1)}{2n-1}} \rightarrow 0
\end{displaymath}
and these constraints are reduced to 
\begin{equation}\label{44}
\frac{1-n}{2n-1}Z^{-1} Z^{;\nu}\sigma_{\alpha\beta\mu\nu}+\sigma_{\alpha\beta\mu\nu}{}^{;\nu}=0.
\end{equation}
\item Finally, eqs. (41) as the propagation ones, are reduced in these limits to 
\begin{equation}\label{n45}
\pi_{\mu\nu;\alpha}{}^{;\alpha}-4\sigma_{\alpha\mu\nu\beta}{}^{;\alpha\beta}
-\frac{1}{\lambda^2}\pi_{\mu\nu}=0.
\end{equation}
The fields $\pi$ and $\sigma$ are subject to only nine equations (45) and to the constraints $\pi_{\mu\nu}{}^{;\nu}=0$ 
and (44).
\end{enumerate}

We now check if in this limit, where all gravitational couplings have been removed, the two fields can exist independently. 
If we set $\sigma_{\alpha\mu\nu\beta}=0$, constraints (44) vanish identically. The field $\pi_{\mu\nu}$ 
is subject to the standard tensorial Klein--Gordon equations and to the same constraints as in the case 
where the term $f(W)$ is absent in the Lagrangian: hence, it has the same properties.
If we instead set $\pi^{\mu\nu}=0$, the tensorial momentum is subject to 
\begin{equation}\label{n46}
\sigma_{\alpha\mu\nu\beta}{}^{;\alpha\beta}=0
\end{equation}
and to the constraints (44). It is well known \cite{Ie} that if a massless field $H_{\alpha\mu\nu\beta}$ has all 
algebraic symmetries of the conformal curvature tensor and satisfies linear field equations 
$H_{\alpha\mu\nu\beta}{}^{;\beta}=0$, then it has spin two. Therefore, eq.(46) would be compatible with 
$\sigma_{\alpha\mu\nu\beta}$ being a massless spin--two field, however  equations (46) and (44) are inconsistent. 
In fact, take divergence $\nabla_{\alpha}$ of (44) and symmetrize the resulting equation in $\beta\mu$, then (46) 
may be applied. One arrives at a system of 9 nonlinear second order equations,
\begin{equation}\label{n47}
(Z\,Z_{;\alpha\beta}-\frac{n}{2n-1}Z_{;\alpha}Z_{;\beta})\,\sigma^{\alpha\mu\nu\beta}=0.
\end{equation}
These depend on the power $n$ in $f(W)$ whereas solutions of (46) are $n$--independent. 
Consistency of solutions to (47) with those to (46) requires the former be 
$n$--independent and this implies that the $n$--dependent term in (47) must vanish,
\begin{equation}\label{n48}
Z_{;\alpha}Z_{;\beta}\,\sigma^{\alpha\mu\nu\beta}=0,
\end{equation}
then (47) is reduced to 
\begin{equation}\label{n49}
Z_{;\alpha\beta}\,\sigma^{\alpha\mu\nu\beta}=0.
\end{equation}
Furthermore, multiplying (44) by $Z^{;\alpha}$ and applying (48) one gets a new constraint, 
$Z^{;\alpha} \sigma_{\alpha\beta\mu\nu}{}^{;\nu}=0$. This constraint, together with the constraint 
(48) and equations (46) and (49), form a system which is consistent only for 
$\sigma^{\alpha\mu\nu\beta}=0$. Clearly this inconsistency is due to truncating eqs. (41) and (43) and 
invalidating eq. (35). The inconsistency shows that the momentum $\sigma_{\alpha}{}^{\beta\mu\nu}$, 
unlike the momentum $\pi^{\mu\nu}$, does not exist as an autonomous physical field in flat spacetime. It 
makes sense only as an auxiliary notion derived from the conformal curvature.\\

We conclude this section by stating that the description of the gravitational system (31) 
in terms of the momenta $\pi^{\mu\nu}$ and $\sigma_{\alpha}{}^{\beta\mu\nu}$ is defective. 
There are too few propagation equations for the momenta and the definition of $\sigma$ gives 
rise to constraints which exclude nonzero $\sigma$ in flat spacetime. In other terms NLG theory (31) 
is not equivalent to Einstein's gravity generated by fields $\pi^{\mu\nu}$ and 
$\sigma_{\alpha}{}^{\beta\mu\nu}$ carrying real masses and definite spins. Degrees of freedom 
associated with $f(W)$, if there are any, cannot be expressed by $\sigma_{\alpha}{}^{\beta\mu\nu}$.

\section{A scalar momentum conjugate to the Weyl tensor}
The other way of dealing with the Weyl tensor contribution to (31) is to assume that it 
carries one independent degree of freedom described by a 
scalar conjugate momentum. One introduces the momenta $\chi$ and 
$\pi^{\mu\nu}$ as previously, according to (10) and (11), and 
\begin{equation}\label{n50}
\sigma \equiv \frac{\partial L}{\partial W}=\frac{1}{k}f'(W).
\end{equation}
Dimensionality of $\sigma$ is $[\sigma]=[\kappa R/W]=ML^2T^{-1}$. The inverse transformation 
to (50) is $W=v(\sigma)$ and actually $v(\sigma)$ is a function of the product 
$k\sigma$. We stress that this procedure, that we have already outlined in Sect.~3, is beyond the setup of the generalized 
Legendre transformation described in \cite{MFF2}. Here, we expect that one of the two scalar fields, $\chi$ and $\sigma$,  
has no independent dynamics, but we need to assess whether the other scalar field can become physically independent from 
$\pi^{\mu\nu}$ and thus provide the 8th DOF.

Again, to be able to perform explicit calculations we adopt the ansatz (32) for $f(W)$; then 
\begin{equation}\label{n51}
 W=v(\sigma)=(k\sigma)^{\frac{1}{n-1}}. 
\end{equation}

The scalar function $H$ is 
\begin{eqnarray}\label{n52}
&
H \equiv \frac{\partial L}{\partial R}(\chi) R(\chi) +
\frac{\partial L}{\partial S_{\alpha\beta}}(\pi) S_{\alpha\beta}(\pi) + 
\frac{\partial L}{\partial W}(\sigma) W(\sigma) - L(g, \chi, \pi,\sigma)=
& \nonumber\\
&
= 3m^2\chi^2 -\frac{m^2}{4}\pi^{\alpha\beta} \pi_{\alpha\beta} +
\sigma v(\sigma) -\frac{1}{k}f(v(\sigma)) 
&
\end{eqnarray}
and it generates the Helmholtz Lagrangian
\begin{eqnarray}\label{n53}
&
L_H(g, R, \chi, S_{\mu\nu}, \pi, W, \sigma) \equiv  
\frac{\partial L}{\partial R}(\chi) R(g) +
\frac{\partial L}{\partial S_{\alpha\beta}}(\pi) S_{\alpha\beta}(g) + 
\frac{\partial L}{\partial W}(\sigma) W(g) -H=
& \nonumber\\
&
=\kappa R + \chi R + \pi^{\alpha\beta}S_{\alpha\beta} -
3m^2\chi^2 +\frac{m^2}{4}\pi^{\alpha\beta} \pi_{\alpha\beta} +
\sigma W -\sigma v(\sigma) +\frac{1}{k}f(v(\sigma)).
&
\end{eqnarray}
The Lagrange equations for the momenta are (38) and $W=v(\sigma)$, hence once again they are
equivalent to quasi--Einsteinian ones (39) and generate the constraints (40). The metric variation 
of $L_H$ generates a second system of Einstein field equations with the energy--momentum tensor 
being a sum of uniquely defined tensors $t_{\mu\nu}(g,\chi,\pi)$ and $\tau_{\mu\nu}(g,\sigma)$. 
Clearly $t_{\mu\nu}$ is equal to $T_{\mu\nu}$ given in (41) with all the $\sigma$--dependent terms 
discarded. From $g^{\mu\nu} \pi_{\mu\nu}=0$ it follows that $t_{\mu\nu}$ is traceless, 
$g^{\mu\nu} t_{\mu\nu}(g,\chi,\pi)=0$. Yet $\tau_{\mu\nu}$ is defined in the standard way by
\begin{equation}\label{n54}
-\frac{1}{2}\sqrt{-g} \tau_{\mu\nu}(g, \sigma) \equiv 
\frac{\delta}{\delta g^{\mu\nu}}\Big[\sqrt{-g} [\sigma W -\sigma v(\sigma) + 
\frac{1}{k}f(v(\sigma))]\Big],
\end{equation}
is equal to
\begin{equation}\label{n55}
\tau_{\mu\nu}(g, \sigma)= [\frac{1}{k}f(v(\sigma))-\sigma v(\sigma)]g_{\mu\nu} +
 2\sigma C_{\alpha\mu\nu\beta} R^{\alpha\beta} +
4\nabla^{\alpha}\nabla^{\beta} (\sigma C_{\alpha(\mu\nu)\beta})
\end{equation}
and its trace is 
\begin{equation}\label{n56}
 g^{\mu\nu} \tau_{\mu\nu}(g, \sigma)=4 [\frac{1}{k}f(v(\sigma)) - \sigma v(\sigma)].
\end{equation}
This trace cannot identically vanish, otherwise employing $\frac{df}{dv}=k\sigma$ one 
gets a differential equation for $f$, $v\,\frac{df}{dv}-f(v)=0$, having the unique 
solution $f(v)=Cv$, contrary to the assumption $f''(v)\neq 0$. For (32) the trace is 
\begin{equation}\label{n57}
\tau_{\mu\nu} g^{\mu\nu} = 4 (\frac{1}{n}-1)(k\sigma^n)^{\frac{1}{n-1}},
\end{equation}
requiring $n>1$. For solutions one replaces $G_{\mu\nu}$ in $8\pi G(t_{\mu\nu}+
\tau_{\mu\nu})=G_{\mu\nu}$ according to (39) and gets
\begin{equation}\label{n58}
8\pi G(t_{\mu\nu} +\tau_{\mu\nu})=-\frac{m^2}{2}(\pi_{\mu\nu}+3\chi g_{\mu\nu})
\end{equation}  
and this equality is crucial. Firstly, the trace of (58), upon applying $g^{\mu\nu} 
t_{\mu\nu}=0$ and (56), provides an expression for $\chi$, 
\begin{equation}\label{n59}
\chi = \frac{16\pi G}{3m^2}[\sigma v(\sigma) -\frac{1}{k} f(v(\sigma))].
\end{equation}
The momentum $\chi$ is eliminated as being a function of $\sigma$ analogous to (42). One 
computes from (59) the derivative $\chi_{;\mu}$ and constraints (40) read now 
\begin{equation}\label{n60}
\pi_{\mu\nu}{}^{;\nu}+ \frac{16\pi G}{m^2}v\sigma_{;\mu}=0.
\end{equation}
Secondly, as in the tensorial case, equations (58) serve as a system of propagation equations 
for the doublet $\{\pi_{\mu\nu},\sigma\}$. To this end one recasts the Weyl tensor terms 
in $\tau_{\mu\nu}$. It is well known that the Bianchi identities generate expressions for the 
single and double divergence of the conformal tensor and one replaces in them the Ricci tensor 
in terms of $\pi_{\mu\nu}$ and $\chi(\sigma)$ according to (38) and finally arrives at 
\begin{eqnarray}\label{n61}
\tau_{\mu\nu} & = & m^2[2(\pi_{\mu\nu;\alpha}-\pi_{\alpha(\mu;\nu)}+g_{\alpha(\mu}
\chi_{;\nu)}-g_{\mu\nu}\chi_{;\alpha})\sigma^{;\alpha}+
\nonumber\\
& + & (\pi_{\mu\nu;\alpha}{}^{;\alpha}-\pi^{\alpha}{}_{(\mu;\nu)\alpha}+\chi_{;\mu\nu}-
g_{\mu\nu}\chi^{;\alpha}{}_{;\alpha})\,\sigma-\sigma C_{\alpha\mu\nu\beta}\pi^{\alpha\beta}]+
\nonumber\\
& + & 4C_{\alpha\mu\nu\beta}\sigma^{;\alpha\beta}+\left[\frac{1}{k}f(v)-\sigma v(\sigma)
\right]g_{\mu\nu}.
\end{eqnarray}
Unlike the $\chi(\sigma)$--dependence, the $\pi_{\mu\nu}$--dependence of $\tau_{\mu\nu}$ 
cannot be eliminated and this shows that the interpretation of $\tau_{\mu\nu}$ as an 
energy--momentum tensor for the field $\sigma$ is of limited sense. Next one 
inserts $t_{\mu\nu}$ and (61) into (58) and after a number of longer manipulations 
one arrives at the following system of equations for the complex $\{\pi_{\mu\nu},\sigma\}$:
\begin{eqnarray}\label{n62}
& &
-m^2\kappa(\pi_{\mu\nu}+3\chi g_{\mu\nu})=(m^2\sigma-1)
\left[\pi_{\mu\nu;\alpha}{}^{;\alpha}+4\chi_{;\mu\nu}-g_{\mu\nu}\chi^{;\alpha}{}_{;\alpha}-
\right.
\nonumber\\
& & \left. {} 
-2C_{\alpha\mu\nu\beta}\pi^{\alpha\beta}-2m^2\chi\pi_{\mu\nu}+m^2\pi^{\alpha}_{\mu}
\pi_{\nu\alpha}-\frac{1}{4}m^2\pi^{\alpha\beta}\pi_{\alpha\beta}g_{\mu\nu}\right]+
\nonumber\\
& & {} 
+4C_{\alpha\mu\nu\beta}\sigma^{;\alpha\beta}+ 2m^2\left[\left(\pi_{\mu\nu;\alpha}-
\pi_{\alpha(\mu;\nu)}-g_{\mu\nu}\chi_{;\alpha}\right)\sigma^{;\alpha}+
\sigma_{(;\mu}\chi_{;\nu)}\right]+
\nonumber\\
& & {} +\left[\frac{1}{k}f(v)-\sigma v(\sigma)\right]g_{\mu\nu}.
\end{eqnarray}
The factor $m^2\sigma-1$ is nonzero since otherwise the definition of $\sigma$ 
implies that $f(W)=(k/m^2)W$ and the linear function is excluded. The full system of 
independent equations for the complex consists of:\\
-- 10 equations (39) for the metric,\\
-- 9 equations (62) for $9+1$ momenta $\pi_{\mu\nu}$ and $\sigma$ (the trace of (62) gives  
expression (59) for $\chi$),\\
-- algebraic expression (59) for $\chi$ as a function of $\sigma$; \\
these are equivalent to a system of ten fourth--order equations for $g_{\mu\nu}$ directly derived 
from (31).

To investigate of whether $\pi_{\mu\nu}$ and $\sigma$ may independently live in flat spacetime, 
one assumes, as in section 5, that $G$, $m^2$ and $k$ tend to 0 at the same rate, so that in this 
limit their ratios $m^2\kappa \equiv \lambda^{-2}$ and $m^2/k\equiv \mu$ are finite and positive.\\
1. Tensors $t_{\mu\nu}$ and $\tau_{\mu\nu}$ get decoupled from the curvature for $G=0$  in EFE. 
This implies $G_{\mu\nu}=0$.\\
2. Equations (39) are consistent with $G_{\mu\nu}=0$ if $m^2=0$.\\
3. The Weyl scalar $W$ must be independent of the scalar momentum $\sigma\neq 0$ and for 
$f(W)=W^n/n$ this is possible iff $k=0$ in $W=v(\sigma)$ given in (51). Then the 
spacetime satisfies $R_{\mu\nu}=0$ and $C_{\alpha\beta\mu\nu}C^{\alpha\beta\mu\nu}=0$. This 
comprises Minkowski space, any plane--parallel (p--p) gravitational wave and few other 
spacetimes. \\
4. For $f(W)$ as in (32), RHS of (59) is proportional to a positive power of $k$ and vanishes 
for $k=0$, hence $\chi=0$.\\
In consequence, eqs. (62) are reduced to 
\begin{equation}\label{n63}
\pi_{\mu\nu;\alpha}{}^{;\alpha}-4C_{\alpha\mu\nu\beta}\sigma^{;\alpha\beta}-
2C_{\alpha\mu\nu\beta}\pi^{\alpha\beta}-\frac{1}{\lambda^2}\pi_{\mu\nu}=0;
\end{equation}
these are traceless. The constraints (40) are simplified to $\pi_{\mu\nu}{}^{;\nu}=0$ (since 
$G/m\rightarrow 0$ in (60)). Now the system comprises (63) and the constraints and is 
underdetermined since there are only 9 eqs. (63) for 10 functions $\pi_{\mu\nu}$ and $\sigma$.\\
In flat spacetime (as this case is crucial for the physical interpretation) the scalar 
$\sigma$ disappears from the field equations and it is clear that it has no physical existence. 
Yet $\pi_{\mu\nu}$ is the standard massive spin--2 field. Just for pure curiosity one may study the 
dynamics of $\sigma$ in a plane gravitational wave for $\pi_{\mu\nu}=0$. It turns out that 
$C_{\alpha\mu\nu\beta}\sigma^{;\alpha\beta}=0$ is reduced in this spacetime to one physically 
bizarre propagation equation which does not uniquely determine $\sigma$ for given initial data.\\ 

We conclude that the momentum $\sigma$ cannot be interpreted as a classical counterpart of a 
quantum particle and should not be introduced into the theory.

\section{The tensorial momentum in the case where the Ricci tensor is absent}
The discussion above has shown that the Ricci term contribution to the Lagrangian consumes 
all degrees of freedom available to nongeometric components of a gravitational multiplet leaving 
no space for the Weyl tensor contribution. We therefore investigate now the case where only $R$ 
and the conformal tensor are present in a gravitational Lagrangian and assume
\begin{equation}\label{n64}
L=\kappa R+aR^2+\frac{1}{k}f(W).
\end{equation}
The $aR^2$ term with $a>0$ is necessary on physical grounds \cite{MS1}. We shall also usually assume 
$f(W)=W^n/n$ for integer $n>1$. Again the momentum associated to $f(W)$ may be either tensorial 
or scalar one and in this section we study the tensorial case. One defines $\chi$ and 
$\sigma_{\alpha}{}^{\beta\mu\nu}$ according to (10) and (33) respectively, introduces the scalar $Z$ 
as in section 5, then $Z(W)$ is given in (34) and the latter is inverted to $W=v(Z)$ giving rise to 
(35) for $C^{\alpha\beta\mu\nu}$. The pseudo--Hamiltonian is
\begin{equation}\label{n65}
H(g,\chi,\sigma)=
\frac{1}{4a}\chi^2+\frac{2kZ}{f'(v(Z))} -\frac{1}{k} f(v(Z))
\end{equation}
and the Helmholtz Lagrangian reads
\begin{equation}\label{n66}
L_H =\kappa R(g) + \chi R(g) + \sigma_{\alpha}{}^{\beta\mu\nu} C^{\alpha}{}_{\beta\mu\nu}(g) -
\frac{1}{4a}\chi^2-\frac{2kZ}{f'(v(Z))} +\frac{1}{k} f(v(Z)).
\end{equation}
The Lagrange equations read $\chi=2aR$ and (35) whereas for the metric one finds
\begin{equation}\label{n67}
(\chi+\kappa)G_{\mu\nu}-\chi_{;\mu\nu}+
g_{\mu\nu}\chi_{;\alpha}{}^{;\alpha}-2\sigma_{\alpha\mu\nu\beta}{}^{;\alpha;\beta}-
R^{\alpha\beta}\sigma_{\alpha\mu\nu\beta}+
g_{\mu\nu}\left[\frac{\chi^2}{8a}+V(Z)\right]=0,
\end{equation}
where the potential term is defined as 
\begin{equation}\label{n68}
V(Z)\equiv -\frac{1}{k}v(Z)f'(v)-\frac{1}{2}\frac{d}{dZ}\left[-\frac{2kZ^2}{f'(v(Z))}+
\frac{1}{k} Z f(v(Z))\right]=\frac{n-1}{2n}k^{\frac{1}{2n-1}} Z^{\frac{n}{2n-1}},
\end{equation}
it is significant that $k$ is here in a positive power. One sees a substantial difference to 
the case where the Ricci term is present: there is no analogue to quasi--Einsteinian equations 
(39) and one cannot eliminate $G_{\mu\nu}$ from (67). The trace of eqs. (67) provides a propagation 
equation for $\chi$, 
\begin{equation}\label{n69}
\Box\chi-\frac{\kappa}{6a}\chi+\frac{4}{3}V(Z)=0;
\end{equation}
$\chi$ is an independent variable and also cannot be eliminated. The triplet $\{g_{\mu\nu}, \chi, 
\sigma_{\alpha\beta\mu\nu}\}$ is subject to $\chi=2aR$, ten eqs. (35), now interpreted as 
propagation equations for the metric with field $\sigma_{\alpha}{}^{\beta\mu\nu}$ as a source, 
nine eqs. (67) and one equation (69) for $\chi$. There are no constraints.

To establish whether $\chi$ and $\sigma$ are classical counterparts of quantum particles one 
considers the case of test fields by putting $G=0=k$ and $a^{-1}=0$. The first three steps in the 
sequence are the same as in section 5 and one gets $R_{\alpha\beta\mu\nu}=0$ and $V(Z)=0$. Then assuming 
that both $G$ and $1/a$ vanish at the same rate, so that $\kappa/a\rightarrow \lambda^{-2}>0$, one 
finds that (69) is simplified to 
\begin{equation}\label{n70}
\Box\chi-\frac{1}{6\lambda^2}\chi=0
\end{equation}
and $\chi$ is the standard massive scalar field. Yet the lack of equations analogous to (39) causes 
that in the flat space limit equations (67) do not form equations of motion for 
$\sigma_{\alpha\mu\nu\beta}$. In a curved spacetime eqs. (67), written in the form of EFE, 
$G_{\mu\nu}=8\pi GT_{\mu\nu}(g,\chi,\sigma)$, define the energy--momentum tensor subject to 
$\nabla^{\nu}T_{\mu\nu}=0$ and these are four equations equivalent to a subset of the full system 
of equations of motion for the two momenta. In flat spacetime the tensor reads (after applying (70)) 
\begin{equation}\label{n71}
T_{\mu\nu}=2\chi_{,\mu\nu}-\frac{1}{3\lambda^2}\chi\eta_{\mu\nu}+4\sigma_{\alpha\mu\nu\beta}{}
^{,\alpha\beta}
\end{equation}
and is subject to $\partial^{\nu}T_{\mu\nu}=0$. It is easy to show that $\partial^{\nu}T_{\mu\nu}$ 
is identically zero and hence gives rise to no equations of motion for $\sigma_{\alpha\mu\nu\beta}$, 
which is completely arbitrary. The description of the system (64) in terms of the tensorial 
momentum conjugate to the Weyl tensor is defective. 

\section{The scalar momentum conjugate to the Weyl invariant if there is no Ricci tensor}
Finally we envisage the scalar momentum conjugate to the Weyl tensor contribution to the 
gravitational system (64). Now one introduces two scalar momenta, $\chi$ as in (10) and $\sigma$ 
according to (50); the inverse to (50) is $W=v(\sigma)$ as in (51). In analogy to (52) and (53) 
one computes
\begin{eqnarray}\label{n72}
H & = & \frac{\chi^2}{4a}+\sigma v(\sigma)-\frac{1}{k}f(v(\sigma)) \qquad \textrm{and}
\nonumber\\
L_H & = & \kappa R(g)+\chi R(g)+\sigma W(g)-\frac{\chi^2}{4a}-\sigma v(\sigma)+
\frac{1}{k}f(v(\sigma)).
\end{eqnarray}
The Lagrange equations comprise $\chi=2aR$, $W(g)=v(\sigma)$ and  
\begin{equation}\label{n73}
\kappa G_{\mu\nu}-\frac{1}{2}(t_{\mu\nu}+\tau_{\mu\nu})=0
\end{equation}
where 
\begin{equation}\label{n74}
t_{\mu\nu}(g,\chi)=-2\chi G_{\mu\nu}+2\chi_{;\mu\nu}-2g_{\mu\nu}\chi_{;\alpha}{}^{\alpha}
+\frac{\chi^2}{4a}g_{\mu\nu}
\end{equation}
and $\tau_{\mu\nu}(g,\sigma)$ is given in (55) and the trace $g^{\mu\nu}\tau_{\mu\nu}\neq 0$ 
is shown in (56) and (57). Tensor $\tau_{\mu\nu}$ involves fourth order derivatives of the 
metric via $C_{\alpha(\mu\nu)\beta}{}^{;\alpha\beta}$ which cannot be expressed in terms of 
derivatives of the momenta. This means that the alleged Einstein field equations (73) are 
inherently of fourth order.

Again the trace of (73) results in a propagation equation for $\chi$, 
\begin{equation}\label{n75}
\chi_{;\mu}{}^{;\mu}-\frac{1}{3a}\chi^2-\frac{\kappa}{6a}\chi+2\left(1-\frac{1}{n}\right)
(k\sigma^n)^{\frac{1}{n-1}}=0;
\end{equation}
it is quasi--linear and contains a source. The full system of equations of motion consists of 
$R=\chi/(2a)$, $W(g)=v(\sigma)$, one eq. (75) for $\chi$ and nine eqs. (73) for 11 components 
of $g_{\mu\nu}$ and $\sigma$; there are no constraints. 

Next one derives the flat spacetime limit for $G\rightarrow 0$, $a^{-1}\rightarrow 0$ and 
$k\rightarrow 0$; this makes sense provided $\frac{\kappa}{a}\rightarrow \lambda^{-2}>0$. 
In this way one arrives at $R=0=G_{\mu\nu}$ and $2W=C_{\alpha\beta\mu\nu} C^{\alpha\beta\mu\nu}=0$, 
as in section 6. The fact that $t_{\mu\nu}(\eta,\chi)$ is linear indicates that both the 
metric and $\chi$ require a redefinition \cite{MS1, S1, S2}. Furthermore, one gets the 
Klein--Gordon equation (70) for $\chi$, whereas for $\sigma$ one may seek for four equations 
arising from $\nabla^{\nu}(t_{\mu\nu}+\tau_{\mu\nu})=0$. In flat spacetime $\tau_{\mu\nu}=0$ 
showing that $\sigma$ carries no energy and has no equation of motion. Yet $\partial^{\nu}
t_{\mu\nu}\equiv 0$ when (70) holds.

\section{Summary and conclusions}
It is natural to expect that gravitational interactions are akin to all 
known physical fields (fundamental fields of the Standard Particle Model) 
which on the quantum level are interpreted as particles with definite 
masses and spins. Even without an explicit quantization, the particle 
interpretation should hold for gravitation. The particle interpretation makes sense 
if each particle species (actually a classical field) may individually 
exist when its gravitational interactions, as well as the interaction with other particle species, 
are switched off (by taking a suitable limit of the coupling constants) in Minkowski space.

In our discussion, we focused on theories of gravity which are envisaged as possible variants to general relativity: 
we write ``variants'', rather than ``alternatives'', to mean that such models share with general relativity the 
fundamental postulates -- in particular, the idea of describing gravitation in terms of metric curvature.
In general relativity, one expects that the ``particle spectrum'' of gravitational interactions reduces to a single 
massless spin--two particle, the graviton, corresponding to excitations of the spacetime metric around the ground state. 
For metric nonlinear gravity theories, where one gets fourth--order equations for the metric, one is led to regard the 
metric itself (with its higher derivatives) as a unifying field, and  gravitation should be described by a multiplet 
of fundamental fields, each with definite mass and spin. According to common wisdom earlier found in the lowest--order 
approximation to a nonlinear Lagrangian, these include, besides the 
graviton field (the physical spacetime metric), a massive scalar field and a massive spin-two field (the latter 
is a ghost, since its kinetic term has negative sign in its quadratic Lagrangian). Here we apply a different method, which is 
exact (no approximations) and again get the graviton and two massive fields, a scalar and a spin-two one (the latter 
may be shown to coincide dynamically, in a linear approximation, with the ghost field mentioned above, though the 
issue is subtle). Thus gravitation in these 
theories is not interpreted as pure curvature, since the gravitational multiplet comprises, besides the metric, 
other fields that should not be viewed as determined by the geometry. Most results in this direction concern 
Lagrangians which depend quadratically on the Ricci tensor. We explored the consequences of a possible Weyl 
tensor dependence, starting from the following assumptions:
\begin{itemize}
\item each of the fundamental fields should carry a definite number of degrees of freedom and should correspond to 
a field which \emph{in flat spacetime, once gravitational interactions are switched off,} can exist independently 
of the other ones (i.e.~can be excited while other fields remain in their ground states); 
\item to obtain (for the full nonlinear model) a proper decomposition of the dynamical variables into a multiplet of fields, 
one should not rely on \emph{ad hoc} tricks: instead, one should rely on a method which can be formulated 
independently of any particular Lagrangian. We identify this method with a 
generalized Legendre transformation, following and extending Kijowski's original proposal \cite{K1} (although 
the version we exploit here is based on the variational formalism, not on the introduction of a symplectic structure). 
Still, for Lagrangians depending on curvature, the application of the method is not unique and one should 
consider different possibilities: the correctness of the choice is verified by counting degrees of freedom for 
both the unifying metric and the multiplet fields.  (Once the multiplet fields have been generated from the 
unifying metric, then they may be subject to further redefinitions which, however, do not alter their 
masses and spins. The freedom of the redefinitions makes any experimental test of the theory ambiguous: 
it is unclear which set -- or ``frame'' -- of multiplet fields is actually measurable; thus, we stress that 
here we consider gravity theories from purely field--theoretical viewpoint and do not take into account observational 
tests);
\item we restrict our investigation to Lagrangians which are reasonably simple: in particular, we consider 
Lagrangians depending on linear and quadratic invariants of the metric curvature. It is known that if the 
Lagrangian depends \emph{linearly} on both the square of the Ricci tensor and the square of the Weyl tensor, 
then the latter can be eliminated by subtracting a full divergence: therefore we allow the square of the Weyl tensor 
to enter the Lagrangian through a nonlinear term, added to the usual generic quadratic Lagrangian.
\end{itemize}
Under these assumptions we have shown that whatever generalized Legendre transformation 
is used to determine the fundamental field associated to the Weyl contribution, 
the resulting field has no independent existence in flat spacetime and carries no degrees of freedom. 

We conclude that within this framework the Weyl tensor does not contribute to the multiplet of gravitational 
fields. Only $R$ and $R_{\mu\nu}$ may contribute and the number of degrees of freedom is either 
seven if the traceless part of Ricci is explicitly present, or three if it is not (and the $R$--dependence is nonlinear). 
The form of the Lagrangian cannot be fully general, since the requirement of forming 
a particle multiplet imposes strict restrictions on both $R$ and $R_{\alpha\beta}$ dependence. 
The mathematically allowed maximal number of 8 DOF is physically unattainable for 
a second--order field--theoretical model satisfying the above assumptions. In other terms, we have shown 
that contrary to the common wisdom mentioned above, a generic nonlinear metric Lagrangian does not describe 
gravity which is equivalent to Einstein's general relativity comprising as a source of the metric 
a number of matter fields (in the sense of being non-geometric quantities) with real masses and definite 
spins. The equivalence only holds for specific Lagrangians and in particular it does not hold if there is 
any explicit Weyl tensor contribution.\\

\textbf{Acknowledgments}. We deeply acknowledge very helpful discussions with Jerzy 
Kijowski, Henryk Arod\'z, Marco Ferraris and Lorenzo Fatibene.\\
The work of L.M.S.  was supported by a grant from the John Templeton Foundation.

\end{document}